\documentclass[floatfix,twocolumn,showpacs,preprintnumbers,amsmath,amssymb,pra,superscriptaddress,longbibliography]{revtex4-1}
\usepackage{color}
\usepackage[usenames,dvipsnames,svgnames,table]{xcolor}
\usepackage[colorlinks=true,linkcolor=blue,urlcolor=blue,citecolor=blue]{hyperref}
\usepackage{mathtools}
\usepackage{graphicx}
\usepackage{dcolumn}
\usepackage{array}
\usepackage{lipsum}
\usepackage{bm}
\usepackage{subfigure}
\usepackage{amssymb}
\usepackage{multirow}
\usepackage{tabularx}
\usepackage{amsmath}
\usepackage{braket}
\usepackage{csquotes}
\graphicspath{{plots/}}
 \usepackage{lipsum}
\usepackage{mathrsfs}
\usepackage{MnSymbol}

\newcommand{\beq}{\begin{equation}}
\newcommand{\eeq}{\end{equation}}
\newcommand{\bea}{\begin{eqnarray}}
\newcommand{\eea}{\end{eqnarray}}

\begin{document}
\title{\emph{Ab initio} Density Response and Local Field Factor of Warm Dense Hydrogen}

\author{Tobias Dornheim}
\email{t.dornheim@hzdr.de}
\affiliation{Center for Advanced Systems Understanding (CASUS), D-02826 G\"orlitz, Germany}
\affiliation{Helmholtz-Zentrum Dresden-Rossendorf (HZDR), D-01328 Dresden, Germany}

\author{Sebastian Schwalbe}
\affiliation{Center for Advanced Systems Understanding (CASUS), D-02826 G\"orlitz, Germany}
\affiliation{Helmholtz-Zentrum Dresden-Rossendorf (HZDR), D-01328 Dresden, Germany}

\author{Panagiotis Tolias}
\affiliation{Space and Plasma Physics, Royal Institute of Technology (KTH), Stockholm, SE-100 44, Sweden}

\author{Maximilian~P.~B\"ohme}
\affiliation{Center for Advanced Systems Understanding (CASUS), D-02826 G\"orlitz, Germany}
\affiliation{Helmholtz-Zentrum Dresden-Rossendorf (HZDR), D-01328 Dresden, Germany}
\affiliation{Technische  Universit\"at  Dresden,  D-01062  Dresden,  Germany}

\author{Zhandos A.~Moldabekov}
\affiliation{Center for Advanced Systems Understanding (CASUS), D-02826 G\"orlitz, Germany}
\affiliation{Helmholtz-Zentrum Dresden-Rossendorf (HZDR), D-01328 Dresden, Germany}

\author{Jan Vorberger}
\affiliation{Helmholtz-Zentrum Dresden-Rossendorf (HZDR), D-01328 Dresden, Germany}

\begin{abstract}
We present quasi-exact \emph{ab initio} path integral Monte Carlo (PIMC) results for the partial static density responses and local field factors of hydrogen in the warm dense matter regime, from solid density conditions to the strongly compressed case. The full dynamic treatment of electrons and protons on the same footing allows us to rigorously quantify both electronic and ionic exchange--correlation effects in the system, and to compare with earlier incomplete models such as the archetypal uniform electron gas [\textit{Phys.~Rev.~Lett.}~\textbf{125}, 235001 (2020)] or electrons in a fixed ion snapshot potential [\textit{Phys.~Rev.~Lett.}~\textbf{129}, 066402 (2022)] that do not take into account the interplay between the two constituents.
The full electronic density response is highly sensitive to electronic localization around the ions, and our results constitute unambiguous predictions for upcoming X-ray Thomson scattering (XRTS) experiments with hydrogen jets and fusion plasmas. All PIMC results are made freely available and can directly be used for a gamut of applications, including inertial confinement fusion calculations and the modelling of dense astrophysical objects. Moreover, they constitute invaluable benchmark data for approximate but computationally less demanding approaches such as density functional theory or PIMC within the fixed-node approximation. 
\end{abstract}

\maketitle

\section{Introduction\label{sec:introduction}}

The properties of hydrogen at extreme temperatures, densities, and pressures are of paramount importance for a wealth of applications~\cite{wdm_book,drake2018high}. An excellent example is given by inertial confinement fusion (ICF)~\cite{Betti2016}, where the fuel capsule (most commonly a deuterium--tritium fuel) has to traverse such warm dense matter (WDM) conditions on its pathway towards ignition~\cite{hu_ICF}. In several recent breakthrough experiments, it has been demonstrated at the National Ignition Facility (NIF)~\cite{Moses_NIF} that it is indeed possible to ignite the fuel~\cite{Zylstra2022} and even produce a net energy gain with respect to the laser energy deposited in the capsule~\cite{Hurricane_RevModPhys_2023,NIF_PRL_2024}.
While undoubtedly a number of challenges remain on the way to an operational fusion power plant~\cite{roadmap}, there is a broad consensus that ICF has great potential to emerge as a future source of safe and sustainable energy. A second class of applications is given by astrophysical objects, such as giant planet interiors~\cite{Militzer_2008,wdm_book,Benuzzi_Mounaix_2014,Helled_NatureReviews_2020} (e.g., Jupiter in our solar system, but also exoplanets) and brown dwarfs~\cite{becker,saumon1}.
Here, the properties of hydrogen are of key relevance to describe the evolution of these objects, and to understand observations such as the gravitational moments of Jupiter~\cite{Nettelmann_2008,Nettelmann_2012,Militzer_2022}.

Despite its apparent simplicity, a rigorous theoretical description of hydrogen remains notoriously elusive over substantial parts of the relevant phase diagram~\cite{Pierleoni_PNAS_2016,Dornheim_review,Filinov_PRE_2023}. A case in point is given by the insulator-to-metal phase transition, which is highly controversial both from a theoretical~\cite{Mazzola2014,Pierleoni_PNAS_2016,Cheng2020,Karasiev_Nature_Comment_2021} and an experimental~\cite{Knudson_Science_2015,Dias_Silvera_Science_2017,Celliers_Science_2018} perspective. The situation is hardly better within the WDM regime that is of particular relevance for both ICF and astrophysical applications~\cite{new_POP,wdm_book}.

Specifically, the WDM regime is usually defined in terms of the Wigner-Seitz radius $r_s=\left(3/4\pi n_e\right)^{1/3}$ (with $n_e=N_e/\Omega$ the electron density) and the degeneracy temperature $\Theta=k_\textnormal{B}T/E_\textnormal{F}$ (with $E_\textnormal{F}$ the Fermi energy of an electron gas at equal density~\cite{quantum_theory}), both of the order of unity~\cite{Ott2018}, i.e., $r_s\sim\Theta\sim1$. Such absence of small parameters necessitates a full treatment of the complex interplay between strong thermal excitations (generally ruling out electronic ground-state methods), Coulomb coupling between both electrons and ions (often ruling out weak-coupling expansions such as Green functions~\cite{stefanucci2013nonequilibrium}), and quantum effects such as Pauli blocking and diffraction (precluding semi-classical schemes such as molecular dynamics simulations with effective quantum potentials~\cite{Bonitz_CPP_2005}).
Consequently, there exists no single method that is reliable over the entire WDM regime~\cite{new_POP}.
In practice, a widely used method is given by a combination of molecular dynamics (MD) simulations of the ions with a thermal density functional theory (DFT)~\cite{Mermin_DFT_1965} treatment of the degenerate electrons based on the Born-Oppenheimer approximation. On the one hand, such DFT-MD simulations are often computationally manageable and, in principle, provide access to various material properties including the equation-of-state~\cite{Holst_PRB_2008,kushal,Danel_PRE_2018,White_PRL_2020}, linear response functions~\cite{Ramakrishna_PRB_2021,Schoerner_PRE_2023,Moldabekov_PRR_2023}, and the electrical conductivity~\cite{Holst_PRB_2011,French_PRE_2022,karasiev_importance,Karasiev_PRB_2022}. On the other hand, the DFT accuracy strongly depends on the employed exchange--correlation (XC) functional~\cite{Clay_PRB_2014,Pierleoni_PNAS_2016}, which has to be supplied as an external input. While the accuracy of many functionals is reasonably well understood in the ground state~\cite{Goerigk_PCCP_2017}, the development of novel, thermal XC-functionals that are suitable for application at extreme temperatures has only started very recently~\cite{ksdt,groth_prl,Karasiev_PRL_2018,Karasiev_PRB_2022,kozlowski2023generalized}.
Moreover, many calculations require additional input, such as the XC-kernel for linear-response time-dependent DFT (LR-TDDFT) calculations of WDM and beyond~\cite{Moldabekov_JCTC_2023}.

Indeed, the linear density response of a system to an external perturbation~\cite{Dornheim_review} constitutes a highly important class of material properties in the context of WDM research. Such linear-response theory (LRT) properties are probed, e.g.,~in X-ray Thomson scattering (XRTS) experiments~\cite{siegfried_review,Froula_2006}, which have emerged as a key diagnostic of WDM~\cite{Dornheim_T_2022,Gregori_PRE_2003}. In principle, the measured XRTS intensity gives one access to the equation-of-state properties of the probed material~\cite{Falk_PRE_2013,Falk_PRL_2014,kraus_xrts}, which is of paramount importance for ICF applications and the modelling of astrophysical objects. Unfortunately, in practice, the interpretation of the XRTS intensity is usually based on uncontrolled approximations such as the widely used Chihara decomposition~\cite{Chihara_1987} into effectively \emph{bound} and \emph{free} electrons. Thus, the quality of inferred parameters such as temperature, density, or ionization remains generally unclear. 

LRT properties are also ubiquitous throughout WDM theory and central to the calculation of stopping power and electronic friction properties~\cite{Moldabekov_PRE_2020,Stopping_2016}, electrical and thermal conductivities~\cite{Veysman_PRE_2016,Jiang2023}, opacity~\cite{Gill_PRE_2021,Hollebon_PRE_2019},  ionization potential depression~\cite{Zan_PRE_2021}, effective ion--ion potentials~\cite{ceperley_potential,Moldabekov_POP_2015}, as well as the construction of advanced nonlocal XC-functionals for thermal DFT simulations~\cite{pribram, Moldabekov_JPCL_2023}.

Owing to its key role, the electronic linear density response has been extensively studied for the simplified uniform electron gas (UEG) model~\cite{review,quantum_theory}, where ions are treated as a rigid homogeneous charge-neutralizing background; see Ref.~\cite{Dornheim_review} and the references therein. These efforts have culminated in different parametrizations of the static local field factor~\cite{dornheim_ML,Dornheim_PRL_2020_ESA,Dornheim_PRB_2021}, which is formally equivalent to the aforementioned XC-kernel -- a basic input for LR-TDDFT simulations and other applications. Very recently, B\"ohme \emph{et al.}~\cite{Bohme_PRL_2022,Bohme_PRE_2023} have extended these efforts and obtained quasi-exact \emph{ab initio} path integral Monte Carlo (PIMC)~\cite{Takahashi_Imada_PIMC_1984,cep,boninsegni1} results for the density response of an electron gas in the external potential of a fixed proton snapshot. On the one hand, these results are directly comparable with thermal DFT simulations of hydrogen, which has given important insights into the accuracy of various XC-functionals~\cite{Moldabekov_PRB_2022,Moldabekov_JCTC_2023,Moldabekov_JCTC_2024}.
On the other hand, these calculations are computationally very inefficient---requiring a large set of independent simulations to estimate the density response at a single wavenumber for a given snapshot---and, more importantly, miss the important dynamic interplay between electrons and protons.

In the present work, we overcome these fundamental limitations by presenting the first \emph{ab initio} PIMC results for the linear density response of full two-component warm dense hydrogen. By treating the electrons and protons dynamically on the same level (i.e., without invoking the Born Oppenheimer approximation), we access all components of the density response, including the electron--electron, electron--proton, and proton--proton local field factors (i.e., XC-kernels). Indeed, a consistent treatment of the interaction between the electrons and ions is crucial to capture the effects of electronic localization around the protons, which has important implications for the interpretation of XRTS experiments. These effects are particularly important for solid state densities ($r_s=3.23$), where they are significant over the entire relevant wavenumber range. They are also substantial for metallic densities ($r_s=2$) and even manifest at strong compression ($r_s=1$) for small wave numbers.

Our simulations are quasi-exact; no fixed-node approximation~\cite{Ceperley1991} is imposed. To deal with the fermion sign problem~\cite{dornheim_sign_problem,troyer} -- an exponential computational bottleneck in quantum Monte Carlo simulations of degenerate Fermi systems -- we average over a large number of Monte Carlo samples, making our simulations computationally very expensive; the cost of this study is estimated to be $\mathcal{O}\left(10^7\right)\,$CPUh. Additionally, we employ the recently introduced $\xi$-extrapolation method~\cite{Dornheim_JCP_2023,Dornheim_JPCL_2024,Xiong_JCP_2022,Dornheim_Science_2024,Dornheim_HBe_2024} to access larger system sizes. We find that finite-size effects are generally negligible at these conditions, which is consistent with previous results for the UEG model~\cite{dornheim_prl,review,Dornheim_JCP_2021}.

We are convinced that our results will open up multiple opportunities for future research. First, the quasi-exact nature of our results makes them a rigorous benchmark for computationally less costly but approximate approaches such as thermal DFT or PIMC within the fixed-node approximation. Moreover, the obtained LRT properties constitute direct predictions for upcoming XRTS experiments with hydrogen jets~\cite{Zastrau,Fletcher_Frontiers_2022} and ICF plasmas~\cite{Moses_NIF}. This makes them also ideally suited to benchmark Chihara models~\cite{Chihara_1987,Gregori_PRE_2003,boehme2023evidence} that are commonly used to diagnose XRTS measurements and infer an equation-of-state. Of particular value are our results for the various partial local field factors, which can be used as input for innumerable applications such as transport property estimates or the construction of novel XC-functionals for thermal DFT simulations. Finally, we note that the presented study of the density response of warm dense hydrogen is interesting in its own right, and gives us new insights into the interplay of electronic localization, quantum effects, and electron--ion correlations in the WDM regime.

The paper is organized as follows: 
In Sec.~\ref{sec:theory}, we provide the relevant theoretical background; a brief introduction of the PIMC method and its estimate of imaginary-time correlation functions (\ref{sec:PIMC}), the full hydrogen Hamiltonian (\ref{sec:Hamiltonian}), the linear density response theory (\ref{sec:density_response}), and its relation to XRTS experiments (\ref{sec:XRTS}). In Sec.~\ref{sec:results}, we present our extensive novel simulation results covering the cases of metallic density (\ref{sec:metallic}), solid state density (\ref{sec:solid_density}), and strong compression (\ref{sec:compressed}). The paper is concluded by a summary and outlook in Sec.~\ref{sec:summary}. Additional technical details are provided in the appendices.

\section{Theory\label{sec:theory}}

\subsection{Hamiltonian\label{sec:Hamiltonian}}

We assume Hartree atomic units (i.e., $\hbar=m_e=e=1$) throughout this work, unless otherwise specified. The full Hamiltonian governing the behaviour of hydrogen is given by
\begin{widetext}
\begin{eqnarray}\label{eq:Hamiltonian}
    \hat{H}_H = - \frac{1}{2}\sum_{l=1}^N\nabla_{l,e}^2 -\frac{1}{2m_p} \sum_{l=1}^{N}\nabla_{l,p}^2+\mathop{\sum_{k=1}^{N}\sum_{l=1}^{N}}_{l<k} W_\textnormal{E}(\hat{r}_l,\hat{r}_k)+\mathop{\sum_{k=1}^{N}\sum_{l=1}^{N}}_{l<k} W_\textnormal{E}(\hat{I}_l,\hat{I}_k)-\sum_{k=1}^N\sum_{l=1}^{N} W_\textnormal{E}(\hat{I}_l,\hat{r}_k)\ ,
\end{eqnarray}
\end{widetext}
where the first and second term correspond to the kinetic energy of the electrons and protons. The pair interaction $W_\textnormal{E}$ is given by the usual Ewald summation, and we follow the conventions introduced in Ref.~\cite{Fraser_PRB_1996}; $\hat{r}$ and $\hat{I}$ denote the position operators of electrons and protons.

\subsection{Path integral Monte Carlo\label{sec:PIMC}}

The \emph{ab initio} PIMC method~\cite{Berne_JCP_1982,Takahashi_Imada_PIMC_1984,cep} constitutes one of the most successful tools for the description of interacting, quantum degenerate many-body systems at finite temperature. The central property is given by the canonical (i.e., inverse temperature $\beta=1/T$, volume $\Omega=L^3$, the simulation box length $L$, and number density $n=N/\Omega$ are fixed) partition function evaluated in the coordinate representation
\begin{widetext}
\begin{eqnarray}\label{eq:Z}
    Z_{N,\Omega,\beta} = \frac{1}{N_\uparrow!N_\downarrow!} \sum_{\sigma_{N_\uparrow}\in S_{N_\uparrow}}\sum_{\sigma_{N_\downarrow}\in S_{N_\downarrow}} \xi^{N_{pp}}  \int\textnormal{d}\mathbf{R} \bra{\mathbf{R}} e^{-\beta\hat{H}_H} \ket{ \hat{\pi}_{\sigma_{N_\uparrow}}\hat{\pi}_{\sigma_{N_\downarrow}}\mathbf{R}}\ , \quad
\end{eqnarray}
\end{widetext}
where the meta-variable $\mathbf{R}=(\mathbf{r}_1,\dots,\mathbf{r}_N,\mathbf{I}_1,\dots,\mathbf{I}_N)^T$ contains the coordinates of both electrons ($\mathbf{r}_l$) and protons ($\mathbf{I}_l$). Throughout this work, we consider the fully unpolarized case with an equal number of spin-up and spin-down electrons, $N_\uparrow = N_\downarrow = N/2$.
Eq.~(\ref{eq:Z}) contains a summation over all possible permutations $\sigma_{N_i}$ of electrons of the same spin orientation $i\in\{\uparrow,\downarrow\}$, which are realized by the corresponding permutation operator $\hat{\pi}_{\sigma_{N_i}}$; $N_{pp}$ denotes the total number of pair permutations of both spin-up and spin-down electrons.
Note that we treat the heavier protons as distinguishable quantum particles (sometimes referred to as \emph{boltzmannons} in the literature), which is exact at the investigated conditions. In fact, even protonic quantum delocalization effects are only of the order of $0.1\%$
here, cf.~Figs.~\ref{fig:H_N14_rs2_theta1_ITCF}, \ref{fig:H_N14_rs3p23_theta1_ITCF}, \ref{fig:H_N14_rs1_theta1_ITCF} below.
The variable $\xi$ determines the type of quantum statistics of the electrons with $\xi=-1$, $\xi=0$, and $\xi=1$ corresponding to Fermi-Dirac, Boltzmann-Maxwell, and Bose-Einstein statistics, respectively.
Only $\xi=-1$ has distinct physical meaning for warm dense hydrogen, although other values can give valuable insights into the importance of quantum degeneracy effects for different observables~\cite{Dornheim_Science_2024,Dornheim_HBe_2024}.

\begin{figure}\centering
\includegraphics[width=0.50\textwidth]{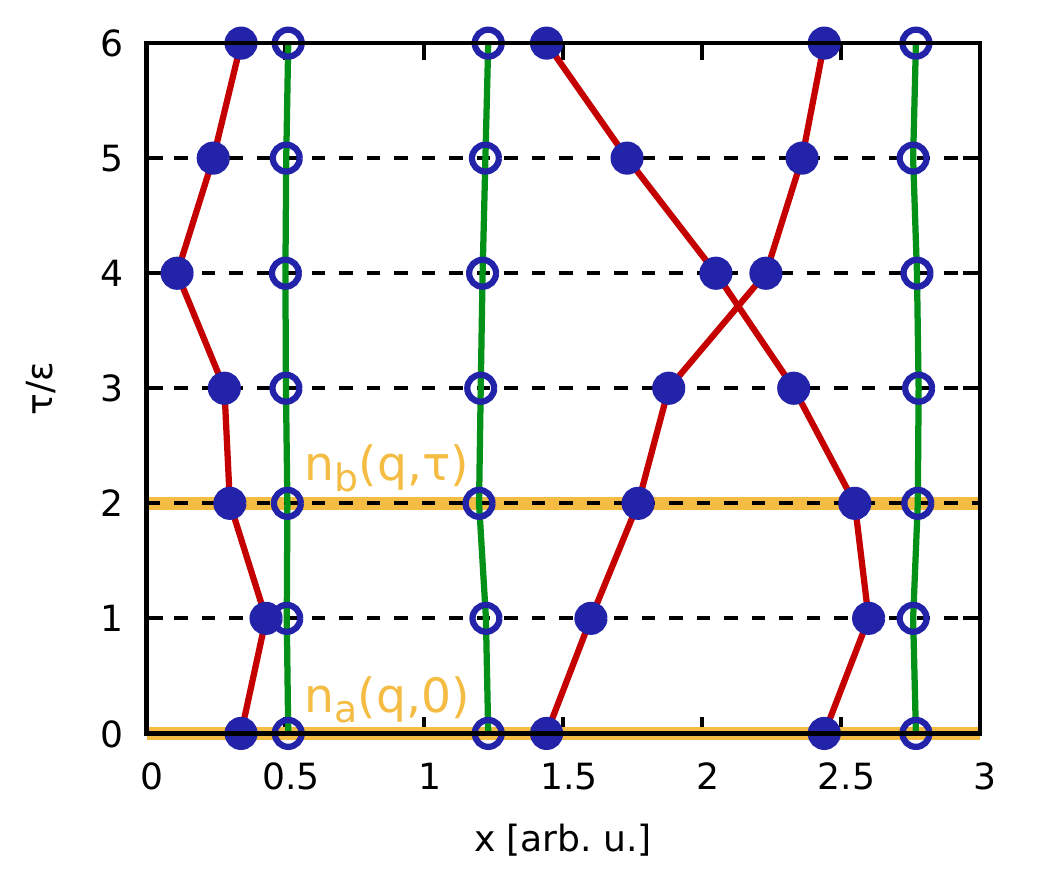}
\caption{\label{fig:schematic} A schematic illustration of the PIMC representation. Shown is a configuration of $N=3$ hydrogen atoms in the $x$-$\tau$-plane for $P=6$ imaginary-time slices. Each electron (filled circles) and each proton (empty circles) is represented by a path along the imaginary-time axis $\tau$ of $P$ coordinates, which are effectively connected by harmonic spring potentials (red and green connections); the extension of the paths is directly related to the respective thermal wavelength $\lambda_T^a=\sqrt{2\pi\beta/m_a}$. The yellow horizontal lines illustrate the evaluation of the density of species $a$ and $b$ in reciprocal space at an imaginary-time distance $\tau$ to estimate the ITCF $F_{ab}(\mathbf{q},\tau)$ [Eq.~(\ref{eq:define_ITCF})].
}
\end{figure} 

A detailed derivation of the PIMC method is beyond our scope and has been presented elsewhere~\cite{cep,boninsegni1,Bohme_PRE_2023}. In essence, the complicated quantum many-body problem of interest, as defined by Eq.~(\ref{eq:Z}), is mapped onto an effectively classical system of interacting ring polymers with $P$ segments each. This is the celebrated \emph{classical isomorphism}~\cite{Chandler_JCP_1981}. A schematic illustration of this idea is presented in Fig.~\ref{fig:schematic}, where we show a configuration of $N=3$ hydrogen atoms in the $\tau$-$x$-plane. In the path integral picture, each particle is represented by an entire path of particle coordinates along the imaginary-time $\tau$ with $P$ imaginary-time slices. The filled and empty circles correspond to the coordinates of electrons and ions, and beads of the same particle on adjacent time slices are connected by a harmonic spring potential. The latter are depicted by the red and green lines for electrons and protons. The extension of these paths corresponds to the thermal wavelength $\lambda_T^a=\sqrt{2\pi\beta/m_a}$, that is smaller by a factor of $1/\sqrt{m_p}\approx0.023$ for the protons.
The basic idea of the PIMC method is to use the Metropolis algorithm~\cite{metropolis} to generate a Markov chain of such path configurations $\mathbf{X}$, where the meta-variable $\mathbf{X}$ contains all coordinates and also the complete information about the sampled permutation structure. A first obstacle is given by the diverging Coulomb attraction, which prevents the straightforward application of the Trotter formula~\cite{kleinert2009path}. To avoid the associated path collapse, we employ the well-known pair approximation~\cite{cep,Militzer_HTDM_2016,Bohme_PRE_2023}, which effectively leads to an off-diagonal quantum potential that remains finite even at zero distance. An additional difficulty arises from the factor $(-1)^{N_{pp}}$ in the case of Fermi statistics. For every change in the permutation structure, the sign of the configuration weight changes. For example, we show a configuration with $N_{pp}=1$ pair exchange in Fig.~\ref{fig:schematic} (i.e., the two electrons on the right), which has a negative configuration weight. In practice, the resulting cancellation of positive and negative terms leads to an exponential increase in the compute time towards low temperatures and large numbers of electrons $N$; this bottleneck is known as the \emph{fermion sign problem} in the literature~\cite{dornheim_sign_problem,troyer}; see also Appendix~\ref{sec:appendix} for additional details. While different, often approximate strategies to mitigate the sign problem have been suggested~\cite{Ceperley1991,Brown_PRL_2013,Dornheim_NJP_2015,Militzer_PRL_2015,Dornheim_JCP_2020,Hirshberg_JCP_2020}, here, we carry out exact direct PIMC solutions that are subject to the full amount of cancellations. The comparably large noise-to-signal level is reduced by averaging over a large number of independent Monte Carlo samples, and our results are exact within the given error bars.

An additional advantage of the direct PIMC method concerns its straightforward access to various imaginary-time correlation functions (ITCF)~\cite{Filinov_PRA_2012,Rabani_PNAS_2002,Dornheim_JCP_ITCF_2021,boninsegni1}. In the context of the present work, the focus lies on the density--density ITCF (hereafter shortly referred to as ITCF)
\begin{eqnarray}\label{eq:define_ITCF}
    F_{ab}(\mathbf{q},\tau) = \braket{\hat{n}_a(\mathbf{q},0)\hat{n}_b(-\mathbf{q},\tau)}\ ,
\end{eqnarray}
which measures the decay of correlations between particles of species $a$ and $b$ along the imaginary-time diffusion process, see yellow lines and symbols in Fig.~\ref{fig:schematic}. As we shall see in Sec.~\ref{sec:density_response}, the ITCF gives direct access to the static density response of warm dense hydrogen from a single simulation of the unperturbed system. The ITCF is interesting in its own right~\cite{Dornheim_MRE_2023,Dornheim_PTR_2022}, and can be measured in XRTS experiments~\cite{Dornheim_T_2022,Dornheim_T2_2022,dornheim2023xray}, see Sec.~\ref{sec:XRTS}.

\subsection{Partial density response functions and partial local field factors\label{sec:density_response}}

The basic idea of the density response formalism is to study a system that is subject to a weak external perturbation which couples linearly to the Hamiltonian~\cite{Dornheim_review}:
\begin{eqnarray}\label{eq:Hamiltonian_perturbed}
    \hat{H}_{\mathbf{q},\omega;A_{e},A_{p}} =  \hat{H}_H &+& 2 A_e \sum_{l=1}^N \textnormal{cos}\left(\mathbf{q}\cdot\hat{\mathbf{r}}_l-\omega t \right) \\\nonumber &+& 2 A_p \sum_{l=1}^N \textnormal{cos}\left(\mathbf{q}\cdot\hat{\mathbf{I}}_l-\omega t \right)\ .
\end{eqnarray}
Here $\hat{H}_H$ denotes the unperturbed Hamiltonian of the full hydrogen system [Eq.~(\ref{eq:Hamiltonian})], and $\mathbf{q}$ and $\omega$ are the wavevector and frequency of the external monochromatic perturbation. It is noted that we distinguish the perturbation amplitude of the electrons ($A_e$) and protons ($A_p$), which allows us to study different species-resolved components of the density response, as we shall discuss in detail. In the limit of an infinitesimally small external perturbation strength, the induced density is a linear function of $A_a$ and linear response theory is applicable~\cite{Dornheim_review,IIT}:
\begin{widetext}
\begin{eqnarray}\label{eq:density}
\delta\braket{\hat{\rho}_a(\mathbf{q},\omega)} = \chi_{aa}^{(0)}(\mathbf{q},\omega)\left(
A_a +\sum_{b}v_{ab}(\mathbf{q})[ 1-G_{ab}(\mathbf{q},\omega)] \delta\braket{\hat{\rho}_b(\mathbf{q},\omega)} 
\right)\ ,
\end{eqnarray}
\end{widetext}
where the induced density perturbation $\delta\braket{\hat{\rho}_a}$ is evaluated in the reciprocal space and where $\chi_{aa}^{(0)}(\mathbf{q},\omega)$ denotes the linear density response of a noninteracting system (of species $a$) at the same conditions. Further, the complete information about species-resolved exchange--correlation effects is contained in the local field factors $G_{ab}(\mathbf{q},\omega)$. Eq.~(\ref{eq:density}) defines a coupled system of equations that explicitly depends on the perturbation $A_a$ of all components.

The fundamental property in linear density response theory is the linear density response function $\chi_{ab}(\mathbf{q},\omega)$ that describes the density response of the species $a$ to a potential energy perturbation applied to the species $b$,
\begin{eqnarray}
    \delta\braket{\hat{\rho}_a(\mathbf{q},\omega)} = A_b \chi_{ab}(\mathbf{q},\omega)\ .
\end{eqnarray}
This \emph{dynamic susceptibility} constitutes a material property of the unperturbed system, and is directly related to the dielectric function~\cite{Hamann_PRB_2020,Hamann_CPP_2020}. Moreover, it is of central importance to LR-TDDFT simulations~\cite{ullrich2012time}, and for the interpretation of XRTS experiments with WDM and beyond, as we explain in Sec.~\ref{sec:XRTS} below.

Let us focus on the static limit $\chi_{ab}(\mathbf{q},0)\equiv\chi_{ab}(\mathbf{q})$ that describes the response to a time-independent cosinusoidal perturbation. Moroni and collaborators~\cite{moroni,moroni2} have presented the first accurate results for the linear static density response function and the local field factor of the ground state UEG based on diffusion QMC simulations. This idea was subsequently adapted to PIMC and configuration PIMC (CPIMC) simulations of the UEG in the WDM regime~\cite{dornheim_pre,groth_jcp}. More specifically, the basic idea is to utilize the formal perturbation strength expansion of non-linear density response theory that reads
\begin{eqnarray}\label{eq:fit}
    \delta\braket{\hat{\rho}_a(\mathbf{q})} = \chi^{(1,1)}_{ab}(\mathbf{q})A_b + \chi^{(1,3)}_{ab}(\mathbf{q})A_b^3  + \mathcal{O}\left(A_b^5\right)\,,
\end{eqnarray}
with $\chi^{(m,l)}_{ab}(\boldsymbol{q})$ the partial static density response of the order $l$ at a harmonic $m$, where $\chi^{(1,1)}_{ab}(\boldsymbol{q})\equiv\chi_{ab}(\boldsymbol{q})$~\cite{Tolias_EPL_2023}. This is a particular case of a general result which states that the total induced density at odd (even) harmonics is given by an infinite series of all $l\geq{m}$ odd (even) powers of the perturbation strength with the coefficients equal to $\chi^{(m,l)}_{ab}(\boldsymbol{q})$~\cite{Tolias_EPL_2023}. Thus, the induced density is estimated for various $A_b$ and a polynomial expansion~\cite{moroni,Dornheim_PRL_2020,Dornheim_review}
\begin{eqnarray}\label{eq:fit1}
    \delta\braket{\hat{\rho}_a(\mathbf{q})} = c_1A_b +c_3A_b^3  + \mathcal{O}\left(A_b^5\right)\ ,
\end{eqnarray}
allows to identify the linear and cubic density response functions at the first harmonic (i.e., at the wavenumber of the perturbation) via the correspondence $c_1=\chi_{ab}(\mathbf{q})$ and $c_3=\chi^{(1,3)}_{ab}(\mathbf{q})$. On the one hand, this \emph{direct perturbation method} is formally exact, and can easily be adapted to other methods such as DFT~\cite{Moldabekov_JCP_2021,Moldabekov_JCTC_2023, Moldabekov_PRB_2023}. On the other hand, it is computationally very expensive, since independent simulations need to be carried out for multiple perturbation amplitudes $A_b$ just to extract the response $\chi_{ab}(\mathbf{q})$ for a single wavevector at a given density and temperature combination. An elegant alternative is given by the imaginary-time version of the fluctuation--dissipation theorem~\cite{Dornheim_MRE_2023}, whose species-resolved version reads
\begin{eqnarray}\label{eq:static_chi}
    \chi_{ab}(\mathbf{q},0) = -\frac{\sqrt{N_a N_b}}{\Omega} \int_0^\beta \textnormal{d}\tau\ F_{ab}(\mathbf{q},\tau)\ ,
\end{eqnarray}
which implies that the species-resolved density responses at all accessible wavevectors can be extracted from a single simulation of the unperturbed system by estimating the corresponding partial $F_{ab}(\mathbf{q},\tau)$. This method has been extensively applied to study the density response of the UEG at finite temperatures over a vast range of parameters~\cite{dornheim_ML,dynamic_folgepaper,dornheim_electron_liquid,dornheim_HEDP,Tolias_JCP_2021,Dornheim_HEDP_2022,Dornheim_PRR_2021,Dornheim_EnergyResponse}.
Here, we employ this second route for the bulk of our new results, while we use the direct perturbation route as an independent cross check.

Let us next consider the local field factors $G_{ab}(\mathbf{q},\omega)$ in more detail. For the UEG, the polarization potential approach leads to the well-known linear density response expression~\cite{kugler1,Tago,review,quantum_theory}
\begin{eqnarray}\label{eq:chi_ee}
    \chi_{ee}(\mathbf{q},\omega) = \frac{\chi^{(0)}_{ee}(\mathbf{q},\omega)}{1 - v_{ee}(q)\left[1 - G_{ee}(\mathbf{q},\omega)\right]\chi^{(0)}_{ee}(\mathbf{q},\omega)}\,.
\end{eqnarray}
$ $\\
Setting $G_{ee}(\mathbf{q},\omega)\equiv0$ in Eq.~(\ref{eq:chi_ee}) leads to the \emph{random phase approximation} (RPA) that describes the electronic density response on a mean-field level. It is pointed out that spin-resolved UEG generalizations have been presented in the literature~\cite{Dornheim_PRR_2022,IIT}, where the spin-up and spin-down electrons are treated as two distinct species.

For multi-component systems, the coupling between the different species makes the situation considerably more complicated. We note that the terms \emph{partial} and \emph{species-resolved} are used interchangeably throughout the text for the associated LRT quantities of multicomponent systems. Introducing the so-called vertex corrected interaction
\begin{eqnarray}
    \theta_{ab}(\boldsymbol{q},\omega)=v_{ab}(\boldsymbol{q})\left[1-G_{ab}(\boldsymbol{q},\omega)\right]\ ,
\end{eqnarray}
the polarization potential approach leads to the following expression for the linear density perturbation $\delta{n}$ that is induced by the potential energy perturbation $\delta{U}$~\cite{kremp_book}
\begin{widetext}
\bea
\delta{n}_{a}(\boldsymbol{q},\omega)=\chi_{aa}^{(0)}(\boldsymbol{q},\omega)\delta{U}_{a}(\boldsymbol{q},\omega)+\displaystyle\sum_{c}\chi_{aa}^{(0)}(\boldsymbol{q},\omega)\theta_{ac}(\boldsymbol{q},\omega)\delta{n}_{c}(\boldsymbol{q},\omega)\,,
\eea
\end{widetext}
which clearly constitutes the generalized version of Eq.~(\ref{eq:density}) for non-monochromatic perturbations. The functional derivative definition of the partial density response function $\chi_{ab}(\boldsymbol{q},\omega)=\delta{n}_{a}(\boldsymbol{q},\omega)/\delta{U}_{b}(\boldsymbol{q},\omega)$ and the identity $\delta{U}_{a}(\boldsymbol{q},\omega)/\delta{U}_{b}(\boldsymbol{q},\omega)=\delta_{ab}$ yield
\begin{widetext}
\bea
\chi_{ab}(\boldsymbol{q},\omega)=\chi_{aa}^{(0)}(\boldsymbol{q},\omega)\delta_{ab}+\displaystyle\sum_{c}\chi_{aa}^{(0)}(\boldsymbol{q},\omega)\theta_{ac}(\boldsymbol{q},\omega)\chi_{cb}(\boldsymbol{q},\omega)\,.
\eea
\end{widetext}
The ideal density response functions $\chi^{(0)}_{ab}$ are nonzero only if $a=b$. Note the reciprocal connections, i.e., 
$\chi_{ab}(\boldsymbol{q},\omega)=\chi_{ba}(\boldsymbol{q},\omega)$, $\theta_{ab}(\boldsymbol{q},\omega)=\theta_{ba}(\boldsymbol{q},\omega)$, $G_{ab}(\boldsymbol{q},\omega)=G_{ba}(\boldsymbol{q},\omega)$, that are a consequence of Newton's third law $v_{ab}(\boldsymbol{q})=v_{ba}(\boldsymbol{q})$. 
The above $3\times3$ set of linear equations can be explicitly solved for the partial density response functions in dependence of $\theta_{ab}(\boldsymbol{q},\omega)$. For hydrogen, $a,b,c=\{e,p\}$, this leads to
\begin{align}
&\chi_{ee}(\boldsymbol{q},\omega)=\displaystyle\frac{1}{\Delta(\boldsymbol{q},\omega)}\chi_{ee}^{(0)}(\boldsymbol{q},\omega)\left[1-\theta_{pp}(\boldsymbol{q},\omega)\chi_{pp}^{(0)}(\boldsymbol{q},\omega)\right]\,,\nonumber\\
&\chi_{pp}(\boldsymbol{q},\omega)=\displaystyle\frac{1}{\Delta(\boldsymbol{q},\omega)}\chi_{pp}^{(0)}(\boldsymbol{q},\omega)\left[1-\theta_{ee}(\boldsymbol{q},\omega)\chi_{ee}^{(0)}(\boldsymbol{q},\omega)\right]\,,\nonumber\\
&\chi_{ep}(\boldsymbol{q},\omega)=\displaystyle\frac{1}{\Delta(\boldsymbol{q},\omega)}\left[\chi_{ee}^{(0)}(\boldsymbol{q},\omega)\theta_{ep}(\boldsymbol{q},\omega)\chi_{pp}^{(0)}(\boldsymbol{q},\omega)\right]\,,\nonumber
\end{align}
where the auxiliary response $\Delta(\boldsymbol{q},\omega)$ is the determinant that is given by
\begin{align}
\Delta(\boldsymbol{q},\omega)&=\left[1-\theta_{ee}(\boldsymbol{q},\omega)\chi_{ee}^{(0)}(\boldsymbol{q},\omega)\right]\left[1-\theta_{pp}(\boldsymbol{q},\omega)\chi_{pp}^{(0)}(\boldsymbol{q},\omega)\right]\nonumber\\&\quad-\theta^2_{ep}(\boldsymbol{q},\omega)\chi_{ee}^{(0)}(\boldsymbol{q},\omega)\chi_{pp}^{(0)}(\boldsymbol{q},\omega)\,\nonumber\,.
\end{align}
When first principle results are available for $\chi_{ab}(\boldsymbol{q},\omega)$, as in our case for the static limit of $\omega=0$, the above $3\times3$ set of linear equations can be explicitly solved for the hydrogen partial local field factors. 
\begin{align}
&\theta_{ee}(\boldsymbol{q},\omega)=\frac{1}{\chi_{ee}^{(0)}(\boldsymbol{q},\omega)}-\frac{\chi_{pp}(\boldsymbol{q},\omega)}{\chi_{ee}(\boldsymbol{q},\omega)\chi_{pp}(\boldsymbol{q},\omega)-\chi^2_{ep}(\boldsymbol{q},\omega)}\,,\label{eq:theta_ee}\\
&\theta_{pp}(\boldsymbol{q},\omega)=\frac{1}{\chi_{pp}^{(0)}(\boldsymbol{q},\omega)}-\frac{\chi_{ee}(\boldsymbol{q},\omega)}{\chi_{ee}(\boldsymbol{q},\omega)\chi_{pp}(\boldsymbol{q},\omega)-\chi^2_{ep}(\boldsymbol{q},\omega)}\,,\label{eq:theta_ii}\\
&\theta_{ep}(\boldsymbol{q},\omega)=\frac{\chi_{ep}(\boldsymbol{q},\omega)}{\chi_{ee}(\boldsymbol{q},\omega)\chi_{pp}(\boldsymbol{q},\omega)-\chi^2_{ep}(\boldsymbol{q},\omega)}\label{eq:theta_ei}\,.
\end{align}
The limit of weak electron--proton coupling is expressed by $\chi_{ee}(\boldsymbol{q},\omega)\chi_{pp}(\boldsymbol{q},\omega)\gg\chi^2_{ep}(\boldsymbol{q},\omega)$ which returns the one-component expressions
\begin{align}
&\theta_{ee}(\boldsymbol{q},\omega)=\frac{1}{\chi_{ee}^{(0)}(\boldsymbol{q},\omega)}-\frac{1}{\chi_{ee}(\boldsymbol{q},\omega)}\,,\nonumber\\
&\theta_{pp}(\boldsymbol{q},\omega)=\frac{1}{\chi_{pp}^{(0)}(\boldsymbol{q},\omega)}-\frac{1}{\chi_{pp}(\boldsymbol{q},\omega)}\,.\nonumber
\end{align}
The opposite limit of strong electron--proton coupling is expressed by $\chi_{ee}(\boldsymbol{q},\omega)\chi_{pp}(\boldsymbol{q},\omega)\ll\chi^2_{ep}(\boldsymbol{q},\omega)$ which again returns the one-component expression but in absence of a non-interacting contribution
\begin{equation}
\theta_{ep}(\boldsymbol{q},\omega)=-\frac{1}{\chi_{ep}(\boldsymbol{q},\omega)}\,.\nonumber
\end{equation}

\subsection{Connection to XRTS experiments\label{sec:XRTS}}

The measured intensity in an XRTS experiment can be expressed as~\cite{Dornheim_T2_2022,Dornheim_review,siegfried_review}
\begin{eqnarray}\label{eq:intensity}
    I(\mathbf{q},\omega) = S_{ee}(\mathbf{q},\omega) \circledast R(\omega)\ ,
\end{eqnarray}
where $S_{ee}(\mathbf{q},\omega)$ is the dynamic structure factor (DSF) and $R(\omega)$ the combined source and instrument function.
In practice, $R(\omega)$ is often known with sufficient accuracy from source monitoring, or from the characterization of a utilized backlighter X-ray source~\cite{MacDonald_POP_2022}.
We note that a direct deconvolution to solve Eq.~(\ref{eq:intensity}) is generally not stable due to noise in the experimental data; a model for $S_{ee}(\mathbf{q},\omega)$ is usually convolved with $R(\omega)$ and compared with $I(\mathbf{q},\omega)$ instead.
The connection between $S_{ee}(\mathbf{q},\omega)$ and the dynamic density response function introduced in the previous section is given by the well-known fluctuation--dissipation theorem~\cite{quantum_theory}
\begin{eqnarray}\label{eq:FDT}
S_{ee}(\mathbf{q},\omega) = - \frac{\textnormal{Im}\{\chi_{ee}(\mathbf{q},\omega)\}}{\pi n_e (1-e^{-\beta\omega})}\,.
\end{eqnarray}
In combination with the Kramers-Kronig causality relation that connects Im$\{\chi_{ee}(\mathbf{q},\omega)\}$ and Re$\{\chi_{ee}(\mathbf{q},\omega)\}$~\cite{quantum_theory}, the fluctuation--dissipation theorem directly implies that $\chi_{ee}(\mathbf{q},\omega)$ and $S_{ee}(\mathbf{q},\omega)$ contain exactly the same information. Moreover, the static density response function $\chi_{ee}(\mathbf{q})$ is related to the DSF via the inverse frequency moment sum rule~\cite{Vitali_PRB_2010}
\begin{eqnarray}\label{eq:inverse_moment}
    \chi_{ee}(\mathbf{q}) = - 2n_e \int_{-\infty}^\infty \textnormal{d}\omega\ \frac{S_{ee}(\mathbf{q},\omega)}{\omega}\ .
\end{eqnarray}
Therefore, accurate knowledge of $\chi_{ee}(\mathbf{q})$ gives direct insights into the low-frequency behaviour of $S_{ee}(\mathbf{q},\omega)$ that is dominated by electron--proton coupling effects such as localization and ionization. Yet, the direct evaluation of the RHS.~of Eq.~(\ref{eq:inverse_moment}) based on experimental XRTS data is generally prevented by the convolution with $R(\omega)$, see Eq.~(\ref{eq:intensity}).
In practice, this problem can be circumvented elegantly by considering the relation between $S_{ee}(\mathbf{q},\omega)$ and the ITCF $F_{ee}(\mathbf{q},\tau)$,
\begin{eqnarray}\label{eq:Laplace}
    F_{ee}(\mathbf{q},\tau) = \mathcal{L}\left[ S_{ee}(\mathbf{q},\omega) \right] = \int_{-\infty}^\infty \textnormal{d}\omega\ S_{ee}(\mathbf{q},\omega)\ e^{-\tau\omega}\ ,
\end{eqnarray}
where $\mathcal{L}[\dots]$ denotes the two-sided Laplace transform operator. Making use of the convolution theorem
\begin{eqnarray}\label{eq:deconvolution}
\mathcal{L}\left[S_{ee}(\mathbf{q},\omega)\right] = \frac{\mathcal{L}\left[S_{ee}(\mathbf{q},\omega)\circledast R(\omega)\right]}{\mathcal{L}\left[R(\omega)\right]}\ ,
\end{eqnarray}
one can thus deconvolve the measured scattering intensity in the imaginary-time domain. In fact, the evaluation of Eq.~(\ref{eq:deconvolution}) turns out to be remarkably stable with respect to the experimental noise~\cite{Dornheim_T2_2022,Dornheim_T_2022,dornheim2023xray} and, thus, gives one direct access to the ITCF from an XRTS measurement. It is then straightforward to obtain the static electron--electron density response function $\chi_{ee}(\mathbf{q})$ from the experimentally measured ITCF via Eq.~(\ref{eq:static_chi}), which amounts to the imaginary time analogue of Eq.~(\ref{eq:inverse_moment}). Our new results for the static density response of full hydrogen thus constitute an unambiguous prediction for upcoming XRTS experiments with hydrogen jets, fusion plasmas, etc.

\begin{figure*}\centering
\hspace*{-0.046\textwidth}\includegraphics[width=0.364\textwidth]{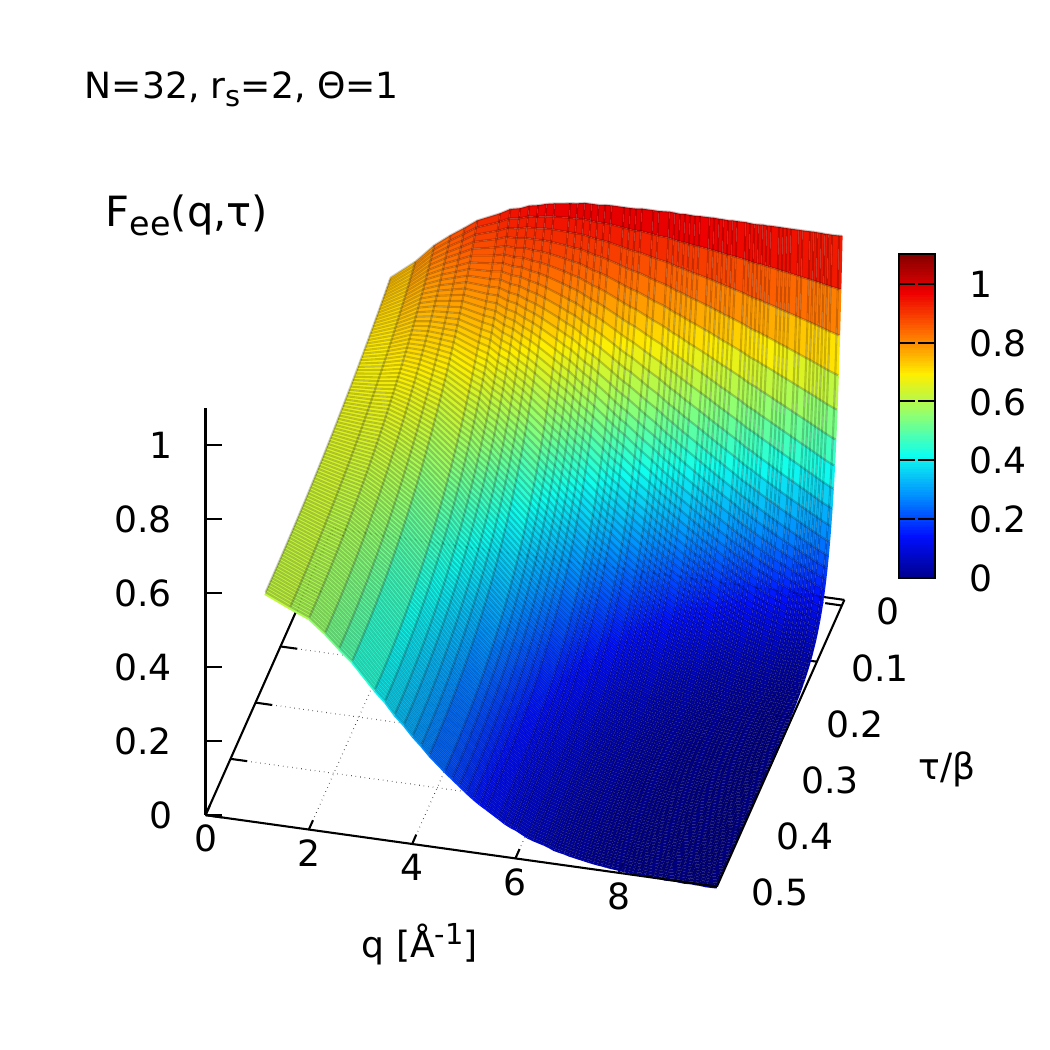}\hspace*{-0.0136\textwidth}\includegraphics[width=0.364\textwidth]{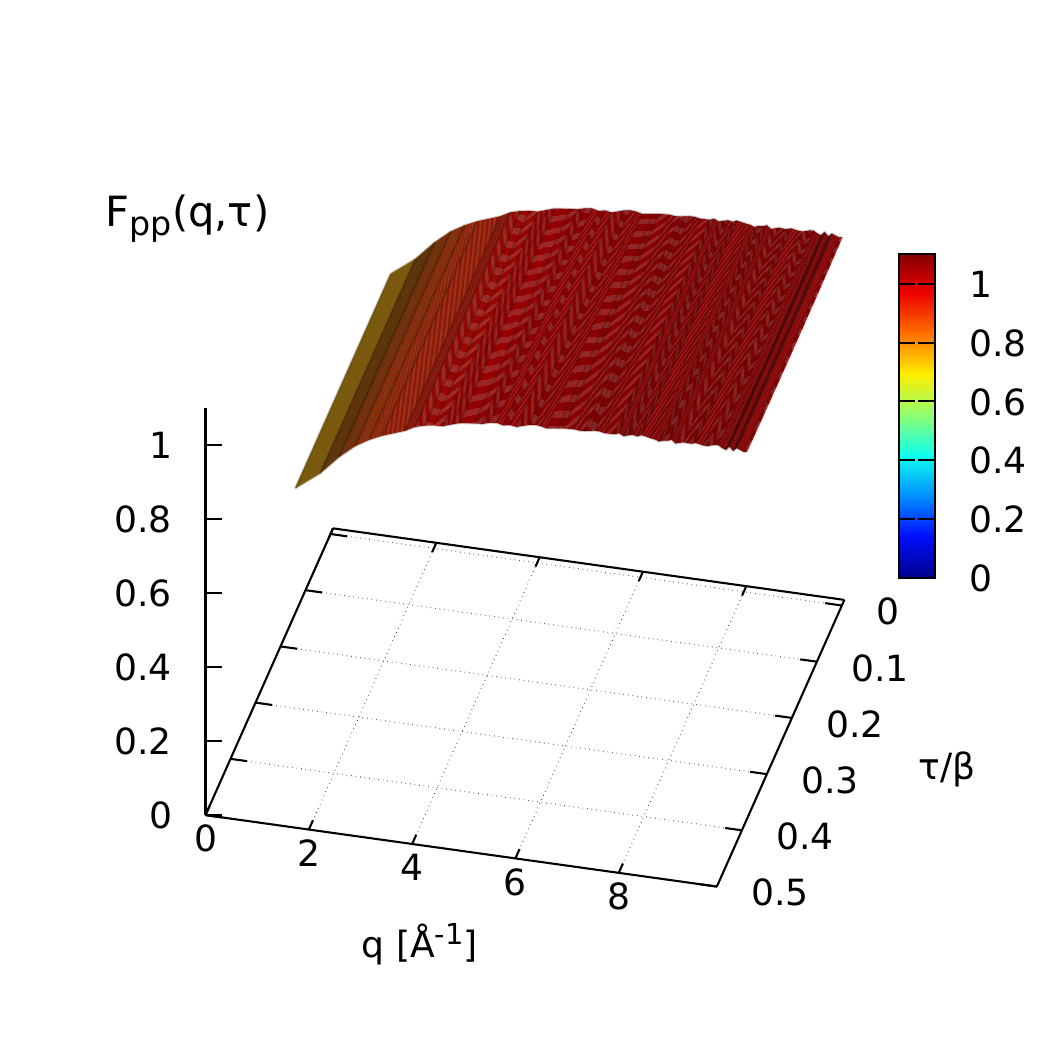}\hspace*{-0.0136\textwidth}\includegraphics[width=0.364\textwidth]{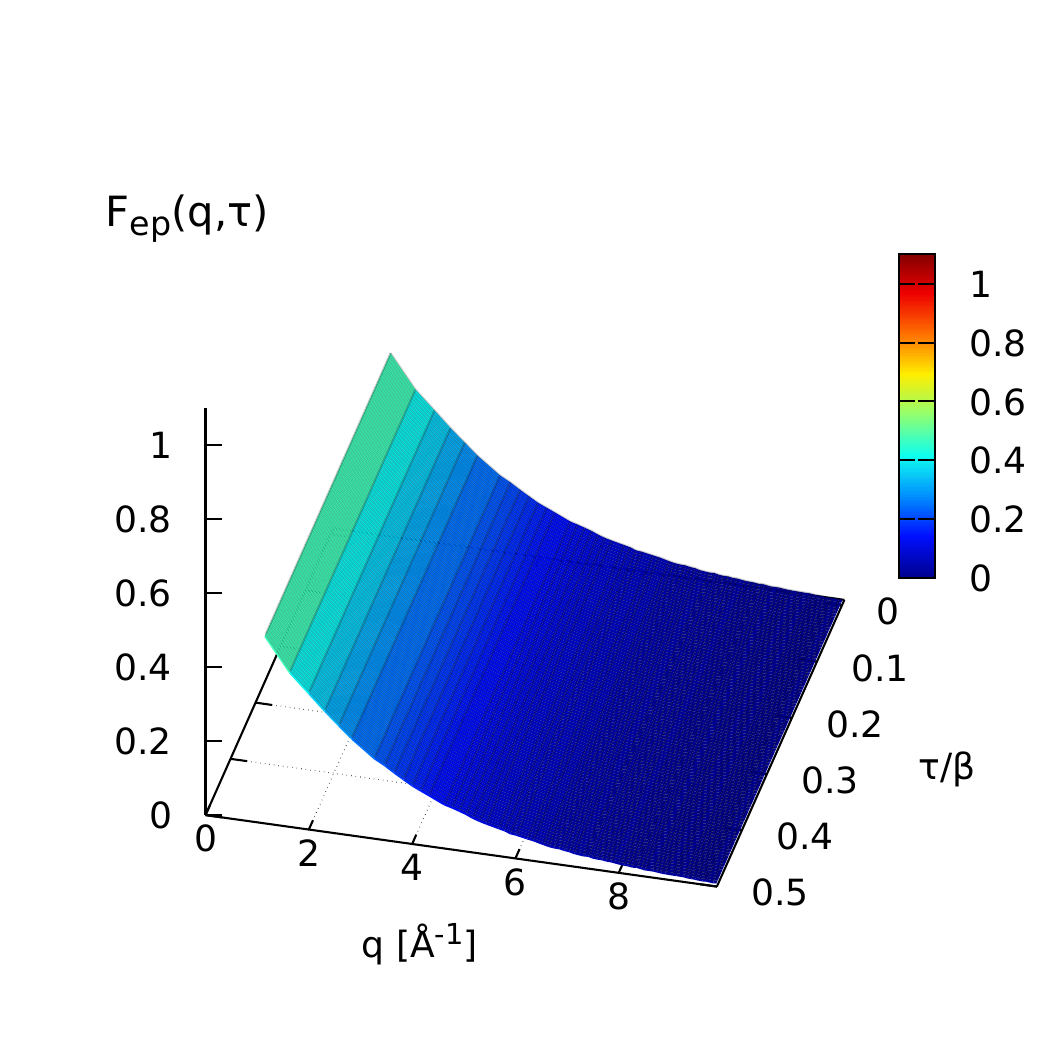}
\caption{\label{fig:ITCF_H_N14_rs2_theta1} \emph{Ab initio} PIMC results for the partial imaginary-time density--density correlation functions of warm dense hydrogen for $N=32$ hydrogen atoms at the electronic Fermi temperature $\Theta=1$ and a metallic density $r_s=2$ in the $\tau$-$q$-plane: electron--electron ITCF $F_{ee}(\mathbf{q},\tau)$ [left], proton--proton ITCF $F_{pp}(\mathbf{q},\tau)$ [center] and electron--proton ITCF $F_{ep}(\mathbf{q},\tau)$ [right].
}
\end{figure*} 

To get additional insights into the physics implications of $\chi_{ee}(\mathbf{q})$, we consider the widely used Chihara decomposition of the DSF~\cite{Chihara_1987,Gregori_PRE_2003,boehme2023evidence},
\begin{eqnarray}\label{eq:Chihara}
    S_{ee}(\mathbf{q},\omega) = S_\textnormal{el}(\mathbf{q},\omega) + \underbrace{S_\textnormal{bf}(\mathbf{q},\omega) + S_\textnormal{ff}(\mathbf{q},\omega)}_{S_\textnormal{inel}(\mathbf{q},\omega)}\ .
\end{eqnarray}
The basic idea of Eq.~(\ref{eq:Chihara}) is to divide the electrons into effectively \emph{bound} and \emph{free} populations. The first contribution to the full DSF is given by the pseudo-elastic component
\begin{eqnarray}\label{eq:elastic}
    S_\textnormal{el}(\mathbf{q},\omega) = W_R(\mathbf{q})\delta(\omega)\ ,
\end{eqnarray}
where the Rayleigh weight contains both the atomic form factor of bound electrons and a screening cloud of free electrons~\cite{Vorberger_PRE_2015}. Therefore, this term originates exclusively from the electronic localization around the protons. The second contribution to the full DSF contains all inelastic contributions, i.e., transitions between bound and free electronic states (and the reverse process, free-bound transitions~\cite{boehme2023evidence}) described by $S_\textnormal{bf}(\mathbf{q},\omega)$ as well as transitions between free states described by $S_\textnormal{ff}(\mathbf{q},\omega)$. It is important to note that the decomposition between \emph{bound} and \emph{free} states is arbitrary in practice and breaks down at high compression where even the orbitals of bound electrons overlap~\cite{Tilo_Nature_2023}; the PIMC method that we use in the present work does not distinguish between bound and free electrons and, as a consequence, is not afflicted by these problems. Nevertheless, the simple picture of the Chihara model Eq.~(\ref{eq:Chihara}) gives important qualitative insights into the density response of the system. For example, Eq.~(\ref{eq:inverse_moment}) directly implies that $\chi_{ee}(\mathbf{q})$ is highly sensitive to electronic localization around the ions, which should result in an increased density response of hydrogen compared to the UEG model~\cite{dornheim_ML,Dornheim_PRL_2020_ESA}; this is indeed what we infer from our PIMC results, as we shall see in Sec.~\ref{sec:results} below.

\section{Results\label{sec:results}}

We use the extended ensemble sampling scheme from Ref.~\cite{Dornheim_PRB_nk_2021} that is implemented in the imaginary-time stochastic high-performance tool for \emph{ab initio} research (\texttt{ISHTAR}) code~\cite{ISHTAR}, which is a canonical adaption of the worm algorithm by Boninsegni \emph{et al.}~\cite{boninsegni1,boninsegni2}. All results are freely available online~\cite{repo} and can be used as input for other calculations, or as a rigorous benchmark for other methods. A short discussion of the $\xi$-extrapolation method~\cite{Xiong_JCP_2022,Dornheim_JCP_2023}, used to simulate larger systems, is given in Appendix~\ref{sec:appendix}.

\subsection{Metallic density: $r_s=2$\label{sec:metallic}}

\begin{figure}\centering
\includegraphics[width=0.48\textwidth]{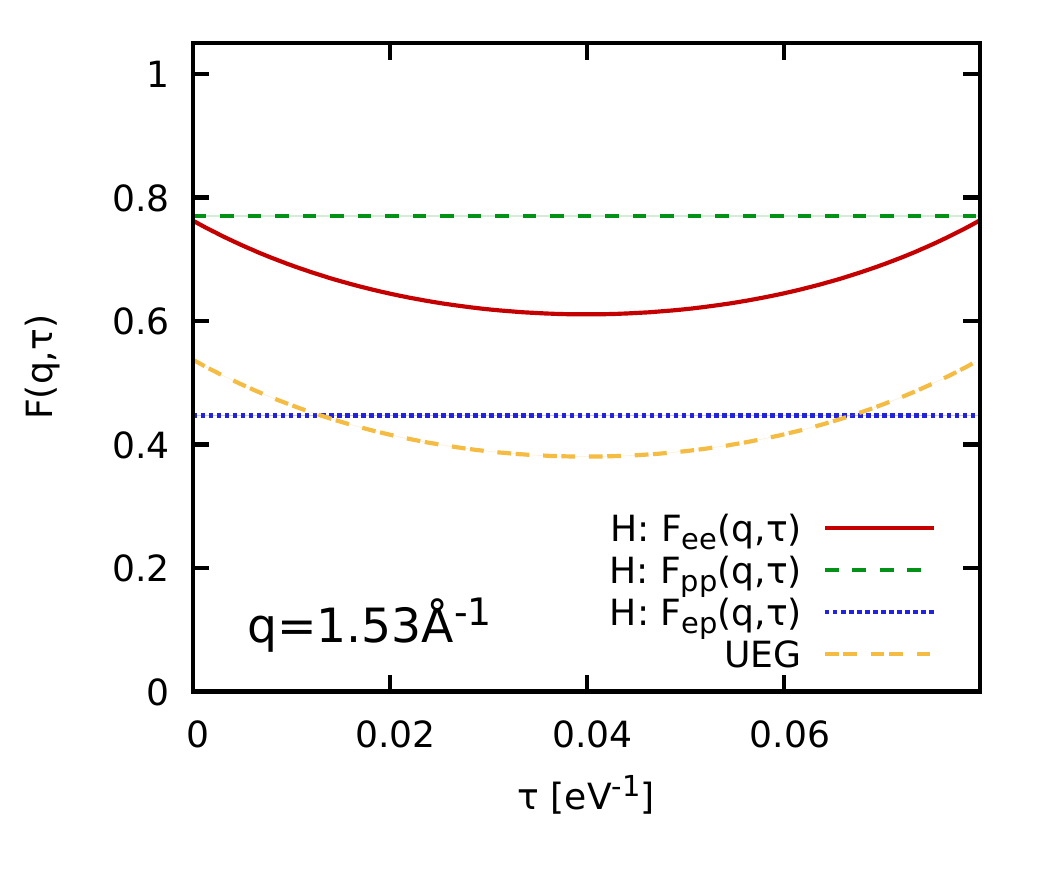}\\\vspace*{-1.3cm}\includegraphics[width=0.48\textwidth]{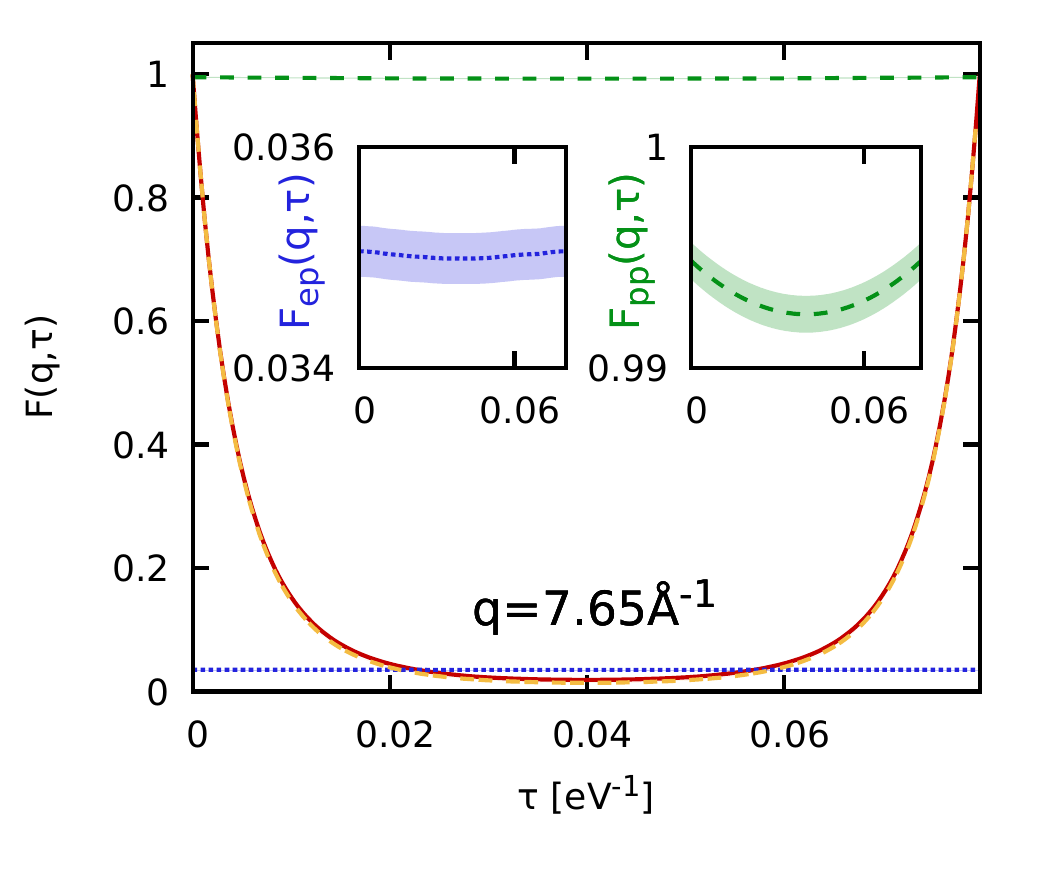}
\caption{\label{fig:H_N14_rs2_theta1_ITCF} \emph{Ab initio} PIMC results for partial hydrogen ITCFs at $r_s=2$ and $\Theta=1$ for $q=1.53\,$\AA$^{-1}$ (or $q=0.84q_\textnormal{F}$) [top] and $q=7.65\,$\AA$^{-1}$ (or $q=4.21q_\textnormal{F}$) [bottom]; solid red: $F_{ee}(\mathbf{q},\tau)$, dashed green: $F_{pp}(\mathbf{q},\tau)$, dotted blue: $F_{ep}(\mathbf{q},\tau)$, double-dashed yellow: UEG model~\cite{Dornheim_MRE_2023}. The shaded intervals correspond to $1\sigma$ error bars. The insets in the right panel show magnified segments around $F_{ep}(\mathbf{q},\tau)$ and $F_{pp}(\mathbf{q},\tau)$.
}
\end{figure}

In this section, we investigate in detail the linear density response of hydrogen at a metallic density $r_s=2$ and at the electronic Fermi temperature $\Theta=1$. We begin our analysis with a study of the ITCF $F_{ab}(\mathbf{q},\tau)$, which constitutes the basis for a significant part of the present work. In Fig.~\ref{fig:ITCF_H_N14_rs2_theta1}, we illustrate $F_{ee}(\mathbf{q},\tau)$ (left), $F_{pp}(\mathbf{q},\tau)$ (center), and $F_{ep}(\mathbf{q},\tau)$ (right) in the relevant $q$-$\tau$-plane. Note that the symmetry relation $F_{ab}(\mathbf{q},\tau)=F_{ab}(\mathbf{q},\beta-\tau)$ holds [see also Fig.~\ref{fig:H_N14_rs2_theta1_ITCF} below], as a consequence of the imaginary-time translation invariance in thermodynamic equilibrium or, equivalently, from the detailed balance relation $S_{ab}(\mathbf{q},\omega) = e^{-\beta\omega}S_{ab}(\mathbf{q},-\omega)$ for the DSF~\cite{Dornheim_T_2022, Dornheim_MRE_2023}. Thus, we can restrict ourselves to the discussion of the interval $\tau\in[0,\beta/2]$. The partial electron--electron ITCF $F_{ee}(\mathbf{q},\tau)$ shown in the left panel exhibits a rich structure that is the combined result of multiple physical effects. The $\tau=0$ limit corresponds to the static structure factor $F_{ab}(\mathbf{q},0)=S_{ab}(\mathbf{q})$, with $S_{ee}(\mathbf{q})$ approaching unity for large wavenumbers, and a finite value for $q\to0$~\cite{Dornheim_HBe_2024}. In addition, $F_{ee}(\mathbf{q},\tau)$ exhibits an increasingly pronounced decay with $\tau$ for large $q$. From a physical perspective, this is due to the Gaussian imaginary-time diffusion process that governs the particle path in the path-integral picture~\cite{Dornheim_MRE_2023,Dornheim_PTR_2022}. In essence, the electrons are quantum delocalized, with the extension of their imaginary-time paths being proportional to the thermal wavelength as it has been explained in the discussion of Fig.~\ref{fig:schematic} above. With increasing $q$, one effectively measures correlations on increasingly small length scales $\lambda=2\pi/q$. For small $\lambda$, any correlations completely decay along the imaginary-time diffusion, and $F_{ee}(\mathbf{q},\beta/2)$ goes to zero. From a mathematical perspective, this increasing $\tau$-decay also follows from the f-sum rule, which states that~\cite{Dornheim_MRE_2023,dornheim2023xray,Dornheim_review,Dornheim_PRB_2023}
\begin{eqnarray}\label{eq:fsum}
   \left. \frac{\partial}{\partial\tau}F_{ab}(\mathbf{q},\tau)\right|_{\tau=0} = -\delta_{ab}\frac{q^2}{2m_a}\ .
\end{eqnarray}
These different trends can also be seen in Fig.~\ref{fig:H_N14_rs2_theta1_ITCF}, where we show the $\tau$-dependence of $F_{ee}(\mathbf{q},\tau)$ for $q=1.53\,$\AA$^{-1}$ (top) and $q=7.65\,$\AA$^{-1}$ (bottom) as the solid red curves. Additional insights come from a comparison with the corresponding UEG results at the same conditions, i.e., the double-dashed yellow curves. For the smaller $q$-value, the DSF of the UEG is dominated by a single broadened plasmon peak in the vicinity of the plasma frequency~\cite{dornheim_dynamic}, leading to a moderate decay with $\tau$. For full hydrogen, we find a very similar $\tau$-dependence, but the entire curve appears to be shifted by a constant off-set; this is a direct signal of the elastic feature [Eq.~(\ref{eq:elastic})] in $S_{ee}(\mathbf{q},\omega)$ that originates from the electronic localization around the protons. For $q=7.65\,$\AA$^{-1}$, on the other hand, the UEG and full hydrogen give very similar results for $F_{ee}(\mathbf{q},\tau)$. In fact, they are indistinguishable around $\tau=0$, as it holds $S_{ee}(\mathbf{q})=1$ in the large-$q$ limit, and both curves exactly fulfill Eq.~(\ref{eq:fsum}). Interestingly, we observe a slightly reduced $\tau$-decay in H compared to the UEG model around $\tau=\beta/2$, which has small though significant implications for $\chi_{ee}(\mathbf{q})$ as we shall see below.

Let us next consider the proton--proton ITCF $F_{pp}(\mathbf{q},\tau)$ and the electron--proton ITCF $F_{ep}(\mathbf{q},\tau)$ which are shown in the center and right panels of Fig.~\ref{fig:ITCF_H_N14_rs2_theta1}. First and foremost, both correlation functions appear to be independent of $\tau$ over the entire depicted $q$-range.
For $F_{ep}(\mathbf{q},\tau)$, this is indeed the case within the given MC error bars, see also the blue curves in Fig.~\ref{fig:H_N14_rs2_theta1_ITCF}. This is consistent with the f-sum rule [Eq.~(\ref{eq:fsum})], which states that $F_{ep}(\mathbf{q},\tau)$ is constant at least in the first order of $\tau$. The proton--proton ITCF exhibits a richer behaviour at a magnified view, see the green curves in Fig.~\ref{fig:H_N14_rs2_theta1_ITCF}. 
Within our PIMC simulations, the protons are treated as delocalized quantum particles just as the electrons, although their thermal wavelength is smaller by a factor of $1/m_p\approx0.02$. This is reflected by the less extended paths along the imaginary-time diffusion process, cf.~Fig.~\ref{fig:schematic} and, consequently, by a strongly reduced $\tau$-decay of $F_{pp}(\mathbf{q},\tau)$ compared to $F_{ee}(\mathbf{q},\tau)$. This is also reflected by the mass in the denominator of the RHS.~of Eq.~(\ref{eq:fsum}).
Remarkably, for large $q$, we can resolve a small, yet significant $\tau$-dependence of $F_{pp}(\mathbf{q},\tau)$ that is of the order of $\sim0.1\%$, see the right inset of the bottom panel of Fig.~\ref{fig:H_N14_rs2_theta1_ITCF}. Overall, it is still clear that the behaviour of both $F_{pp}(\mathbf{q},\tau)$ and $F_{ep}(\mathbf{q},\tau)$ is essentially governed by the corresponding static structure factors $S_{pp}(\mathbf{q})$ and $S_{ep}(\mathbf{q})$, while the $\tau$-dependence is essentially negligible for the linear density response in this regime.

\begin{figure}\centering
\includegraphics[width=0.48\textwidth]{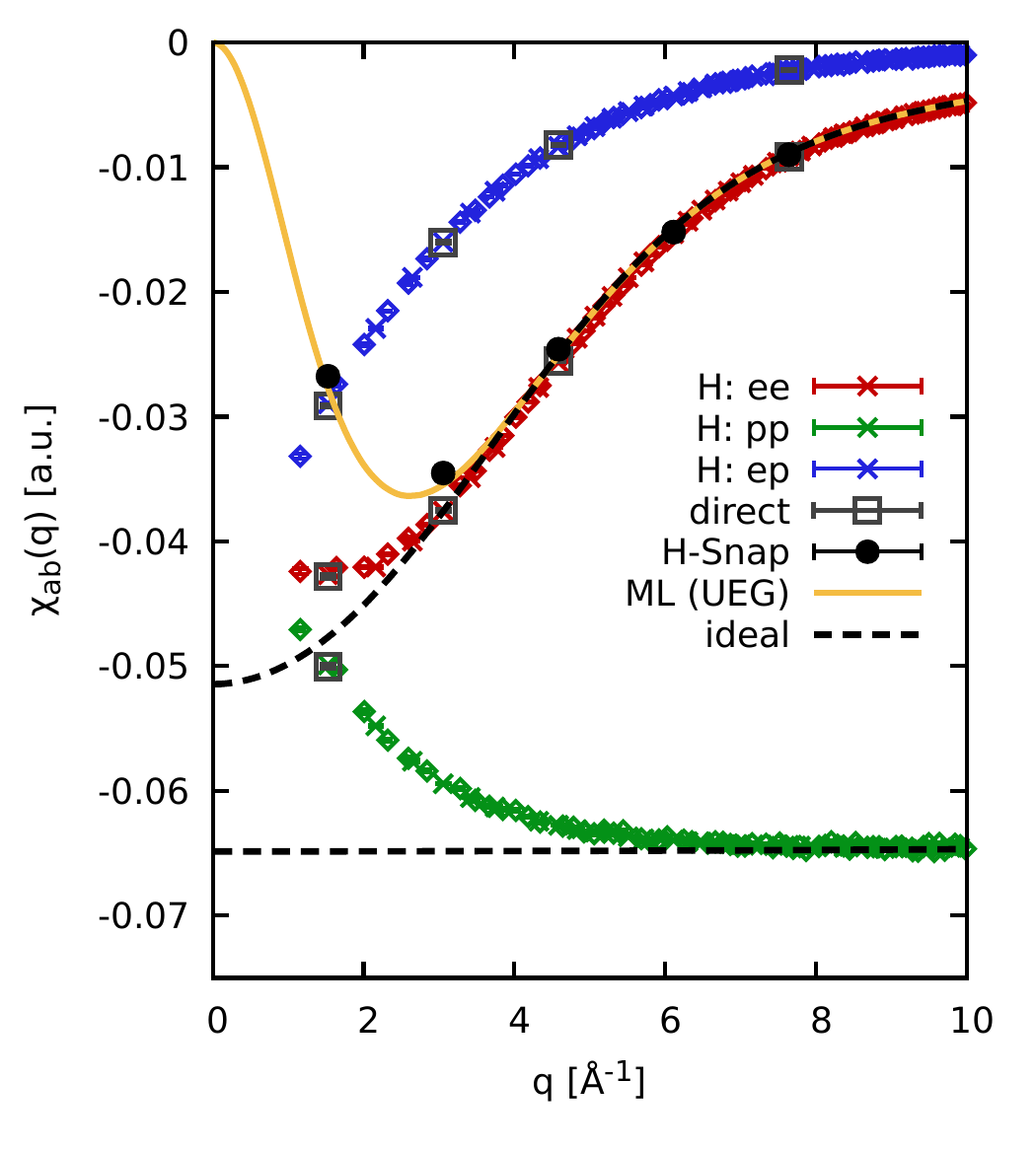}
\caption{\label{fig:H_N14_rs2_theta1_q} \emph{Ab initio} PIMC results for the partial static density responses of hydrogen at $r_s=2$ and $\Theta=1$. Red, green, blue symbols: $\chi_{ee}(\mathbf{q})$, $\chi_{pp}(\mathbf{q})$, $\chi_{ep}(\mathbf{q})$ for full hydrogen evaluated from the ITCF via Eq.~(\ref{eq:static_chi});
grey squares: $\chi_{ee}(\mathbf{q})$, $\chi_{pp}(\mathbf{q})$, $\chi_{ep}(\mathbf{q})$ for full hydrogen evaluated from the direct perturbation approach [Eq.~(\ref{eq:fit})]; black circles: electronic density response of fixed ion snapshot~\cite{Bohme_PRL_2022}; solid yellow line: UEG~\cite{dornheim_ML}; dashed black: ideal density responses $\chi^{(0)}_{ee}(\mathbf{q})$ and $\chi^{(0)}_{pp}(\mathbf{q})$.
}
\end{figure} 

We now proceed to the central topic of the present work, which is the investigation of the partial static density response of warm dense hydrogen. Fig.~\ref{fig:H_N14_rs2_theta1_q} depicts the static density response function $\chi_{ab}(\mathbf{q})$ as a function of the wavenumber $q$. The red, green, and blue symbols have been computed from the ITCF through Eq.~(\ref{eq:static_chi}) for the electron--electron, proton--proton, and electron--proton response, respectively. The crosses and diamonds correspond to $N=14$ and $N=32$ atoms and no dependence on the system size can be resolved. Before comparing the data sets with the other depicted models and calculations, we shall first summarize their main trends. i) the electron--proton density response $\chi_{ep}(\mathbf{q})=\chi_{pe}(\mathbf{q})$ has the same negative sign as $\chi_{ee}(\mathbf{q})$ and $\chi_{pp}(\mathbf{q})$, since the unperturbed protons would follow the induced density of the perturbed electrons, and vice versa. ii) the electron--proton density response $\chi_{ep}(\mathbf{q})$ monotonically decays with $q$ as the electron--proton coupling vanishes in the single-particle limit; the same trend also leads to the well-known decay of the elastic feature in $S_{ee}(\mathbf{q},\omega)$ quantified by the correspondingly vanishing Rayleigh weight $W_\textnormal{R}(\mathbf{q})$. iii) the electron--electron density response $\chi_{ee}(\mathbf{q})$ is relatively flat for $q\lesssim2\,$\AA$^{-1}$ and monotonically decays for larger $q$. Eq.~(\ref{eq:static_chi}) directly implies that this is a quantum delocalization effect. In practice, the static density response is proportional to the area under the corresponding ITCF. The latter vanishes increasingly fast with $\tau$ with increasing $q$ as discussed above, leading to the observed reduction in $\chi_{ee}(\mathbf{q})$. Heuristically, this can be understood as follows. While the static density response of a single (or noninteracting) classical particle is wavenumber independent, the response of a delocalized quantum particle gets reduced when its thermal wavelength is comparable to the perturbation wavelength. In fact, a quantum particle will stop reacting all together when its extension is much larger than the excited wavelength. iv) the proton--proton density response $\chi_{pp}(\mathbf{q})$ increases with $q$ and seemingly becomes constant for $q\gtrsim6\,$\AA$^{-1}$. The reduction in $\chi_{pp}(\mathbf{q})$ for small $q$ is a consequence of the proton--proton coupling (and its interplay with electrons), whereas its behaviour for large $q$ comes from the heavier proton mass and the resulting strongly reduced quantum delocalization. For completeness, note that we can resolve small deviations from the classical limit of $\chi_{pp}^\textnormal{cl}(\mathbf{q})=-n_p\beta$ for large wavenumbers. We also note that the general property $|\chi_{AA}^\textnormal{q}(\mathbf{q})|\leq|\chi_{AA}^\textnormal{cl}(\mathbf{q})|$, valid for static linear response functions associated with any hermitian operator $\hat{A}$~\cite{quantum_theory}, should be respected by all the species-resolved density responses.

\begin{figure}\centering
\includegraphics[width=0.48\textwidth]{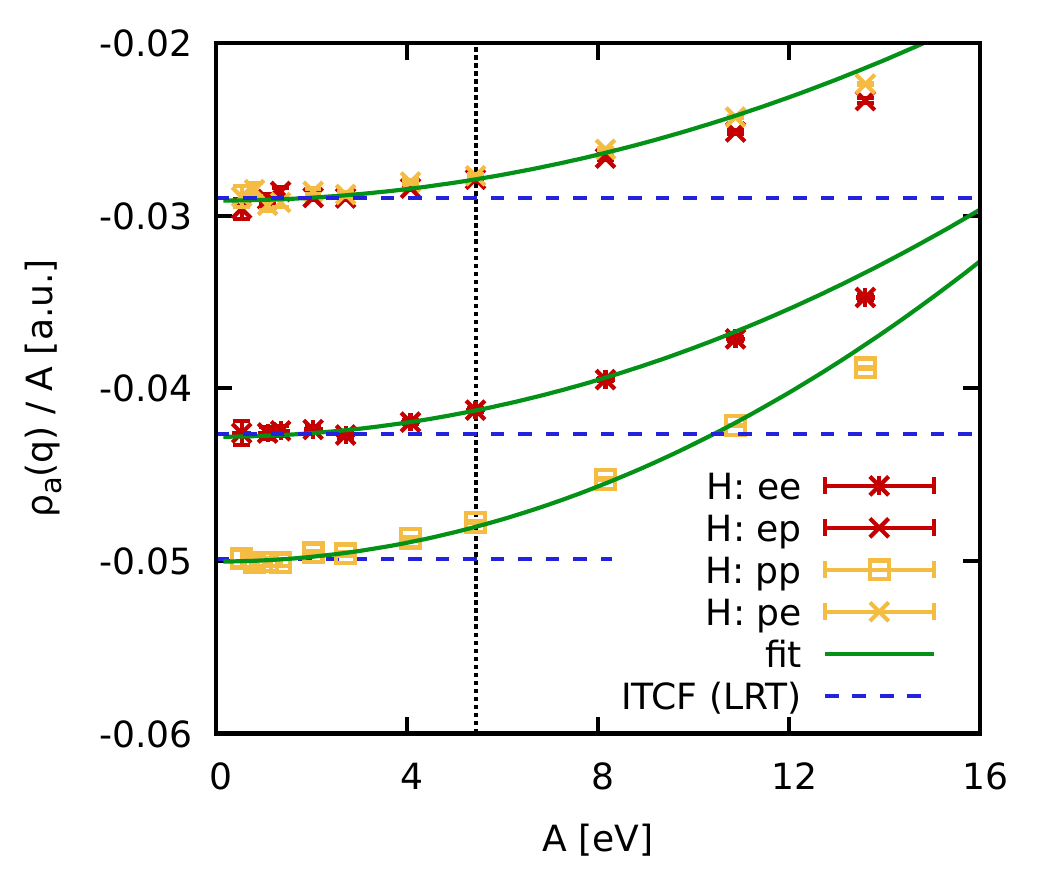}\\\vspace*{-1.cm}\includegraphics[width=0.48\textwidth]{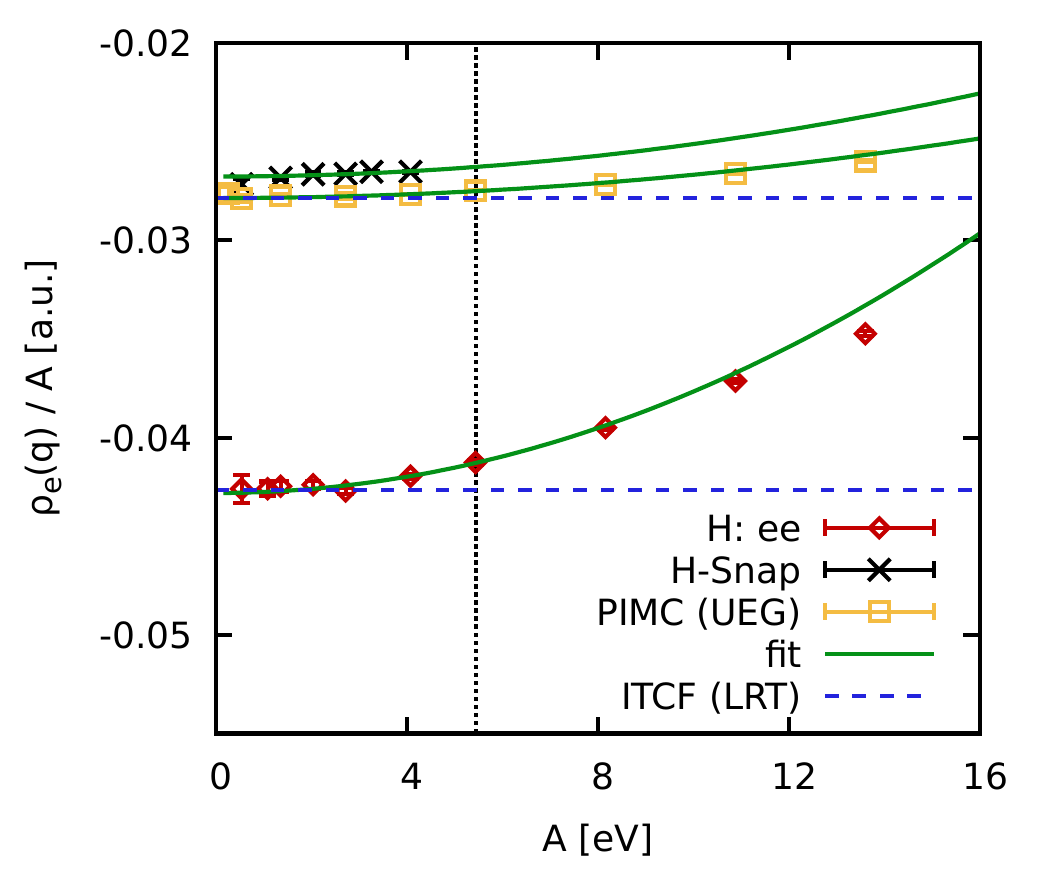}
\caption{\label{fig:H_N14_rs2_theta1_qz1_A} Partial induced density for $q=1.53\,$\AA$^{-1}$ as function of the perturbation strength $A$ [cf.~Eq.~(\ref{eq:Hamiltonian_perturbed})] at $r_s=2$ and $\Theta=1$ for $N=14$ H atoms. The dashed blue lines show the linear-response limit computed from the ITCF via Eq.~(\ref{eq:static_chi}), and the solid green lines the cubic polynomial fits via Eq.~(\ref{eq:fit}). Top: the induced electronic density $\rho_e$ as function of $A_e$ (red stars, $A_p=0$) and $A_p$ (yellow crosses, $A_e=0$); the induced proton density $\rho_p$ as function of $A_p$ (yellow squares, $A_e=0$) and $A_e$ (red crosses, $A_p=0$).
Bottom: the induced electronic density $\rho_e$ as function of the electronic perturbation strength $A_e$ for full hydrogen (red diamonds, $A_p=0$), a fixed proton snapshot~\cite{Bohme_PRL_2022} (black crosses), and the UEG~\cite{Dornheim_PRR_2021} (yellow squares). Additional results are shown in Appendix~\ref{sec:appendix_perturbation}.
}
\end{figure} 

\begin{figure}\centering
\includegraphics[width=0.48\textwidth]{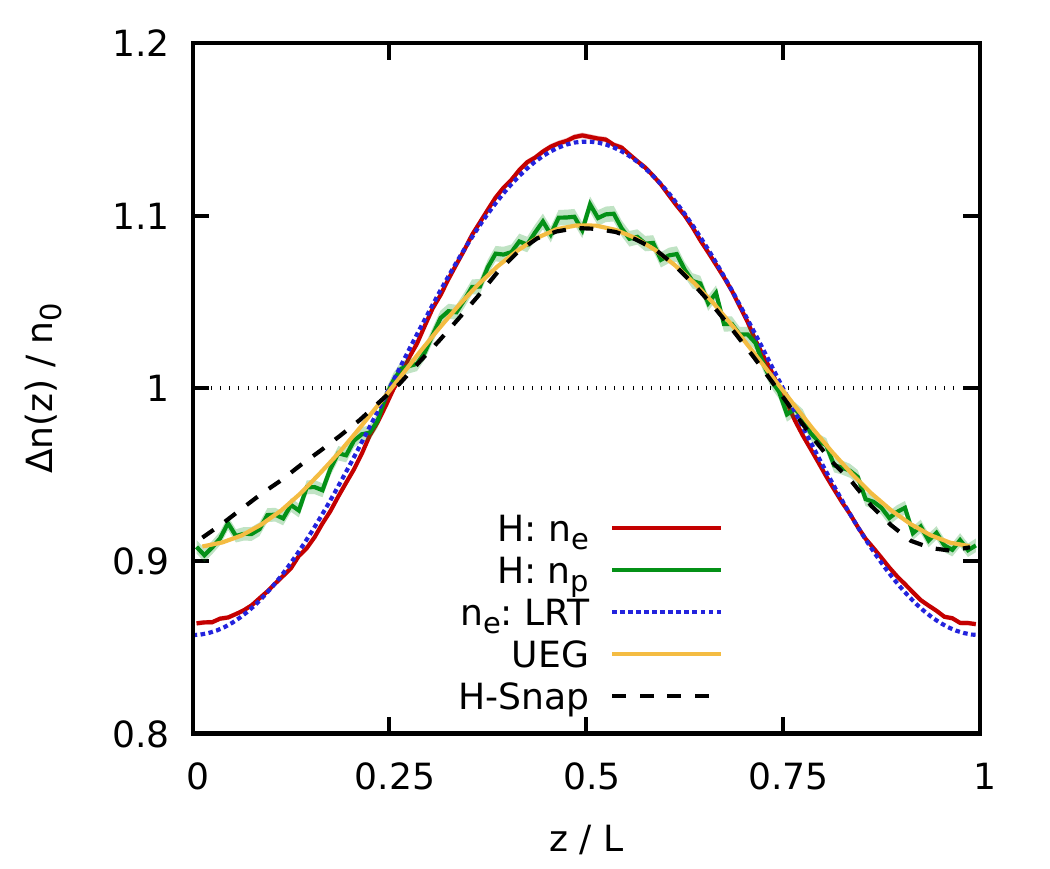}
\caption{\label{fig:strip_rs2_theta1} The induced density change $\Delta n(z)/n_0$ for an electronic perturbation amplitude of  $A_e=0.05\,$Ha ($A_e=1.36\,$eV) [cf.~Eq.~(\ref{eq:Hamiltonian_perturbed})] for $r_s=2$ and $\Theta=1$. Solid red: electron density $n_e(z)$ for full hydrogen (with $A_p=0$); dotted blue: corresponding linear-response prediction [Eq.~(\ref{eq:LRT_density})]; solid green: proton density $n_p(z)$ for full hydrogen (with $A_p=0$); solid yellow: electron density $n_e(z)$ for the UEG model~\cite{Dornheim_PRR_2021}; dashed black: electron density $n_e(z)$ of a fixed proton snapshot~\cite{Bohme_PRL_2022}.
}
\end{figure}

Equally interesting to these observations about the density response of full two-component hydrogen is their comparison to other models and calculations. The grey squares in Fig.~\ref{fig:H_N14_rs2_theta1_q} have been obtained from the \emph{direct perturbation approach}, i.e., from independent PIMC simulations of hydrogen where one component has been perturbed by an external harmonic perturbation, cf.~Eq.~(\ref{eq:Hamiltonian_perturbed}). This procedure is further illustrated in Fig.~\ref{fig:H_N14_rs2_theta1_qz1_A}, where we show the induced density $\rho_a(\mathbf{q})$ as a function of the perturbation amplitude. More specifically, the red crosses in the top panel correspond to the electronic density of full two-component hydrogen induced by the electronic perturbation amplitude $A_e$ (with $A_p=0$) for a wavenumber of $q=1.53\,$\AA$^{-1}$. In the limit of $A_e\to0$, $\rho_e(\mathbf{q})/A_e$ attains a finite value that is given by the static linear density response function $\chi_{ee}(\mathbf{q})$; the latter is shown as the horizontal dashed blue line, as computed from the ITCF $F_{ee}(\mathbf{q},\tau)$ via Eq.~(\ref{eq:static_chi}), and it is in excellent agreement with the red diamonds in the limit of small $A_e$. The solid green lines show cubic fits via Eq.~(\ref{eq:fit}). Evidently, the latter nicely reproduce the $A_e$ dependence of the PIMC data for moderate perturbations, and the linear coefficient then corresponds to the linear density response function; taking into account the deviations between data and fits for $A\gtrsim10\,$eV is possible by including higher-order terms in Eq.~(\ref{eq:fit})~\cite{Tolias_EPL_2023}, which will be pursued in future works. The red crosses have been obtained from the same set of simulations (i.e., $A_e>0$ and $A_p=0$) and depict the corresponding induced density of the protons that is described by $\chi_{ep}(\mathbf{q})$. The unperturbed protons thus do indeed follow the perturbed electrons, although with a somewhat reduced magnitude.
This can be discerned particularly well in Fig.~\ref{fig:strip_rs2_theta1}, where we show the density in real space along the direction of the perturbation for a comparably small electronic perturbation amplitude of $A_e=1.36\,$eV. The solid red line shows the PIMC results for the electronic density and the dotted blue curve the linear-response theory estimate~\cite{Dornheim_PRR_2021}
\begin{eqnarray}\label{eq:LRT_density}
    n_e(\mathbf{r}) = n_0 + 2 A_e\ \textnormal{cos}\left( \mathbf{q}\cdot \mathbf{r}
    \right)\ \chi_{ee}(\mathbf{q})\ ,
\end{eqnarray}
using the ITCF based result for $\chi_{ee}(\mathbf{q})$. Both curves are in excellent agreement, which implies that LRT is accurate in this regime. The green curve shows the proton density from the same calculation. It exhibits the same cosinusoidal form, but with a reduced magnitude. Let us postpone a discussion of the other curves in Fig.~\ref{fig:strip_rs2_theta1} and return to the top panel of Fig.~\ref{fig:H_N14_rs2_theta1_qz1_A}. The yellow squares show results for the induced proton density $\rho_p$ that have been obtained from a second, independent set of PIMC calculations with a finite proton perturbation amplitude $A_p>0$, but unperturbed electrons, $A_e=0$. We find excellent agreement with the ITCF based result for the LRT limit of $\chi_{pp}(\mathbf{q})$, and the cubic fit agrees well with the data. The protons react more strongly to an external static perturbation compared to the electrons, see the discussion of Fig.~\ref{fig:H_N14_rs2_theta1_q} above. Finally, the yellow crosses show the induced electron density $\rho_e$ from the same calculation. We recover the expected symmetry $\chi_{ep}(\mathbf{q})=\chi_{pe}(\mathbf{q})$ within the linear response limit. Interestingly, this breaks down for larger perturbation amplitudes, where the protons again react more strongly than the electrons. This nonlinear effect deserves further explanation, and will be investigated in detail in a dedicated future work.

In the appendix~\ref{sec:appendix_perturbation}, we show more results for the direct perturbation approach for different wavenumbers, which have been employed to obtain the linear density response functions that are shown as the empty squares in Fig.~\ref{fig:H_N14_rs2_theta1_q}. We find perfect agreement with the ITCF based data sets everywhere.

\begin{figure}\centering
\includegraphics[width=0.48\textwidth]{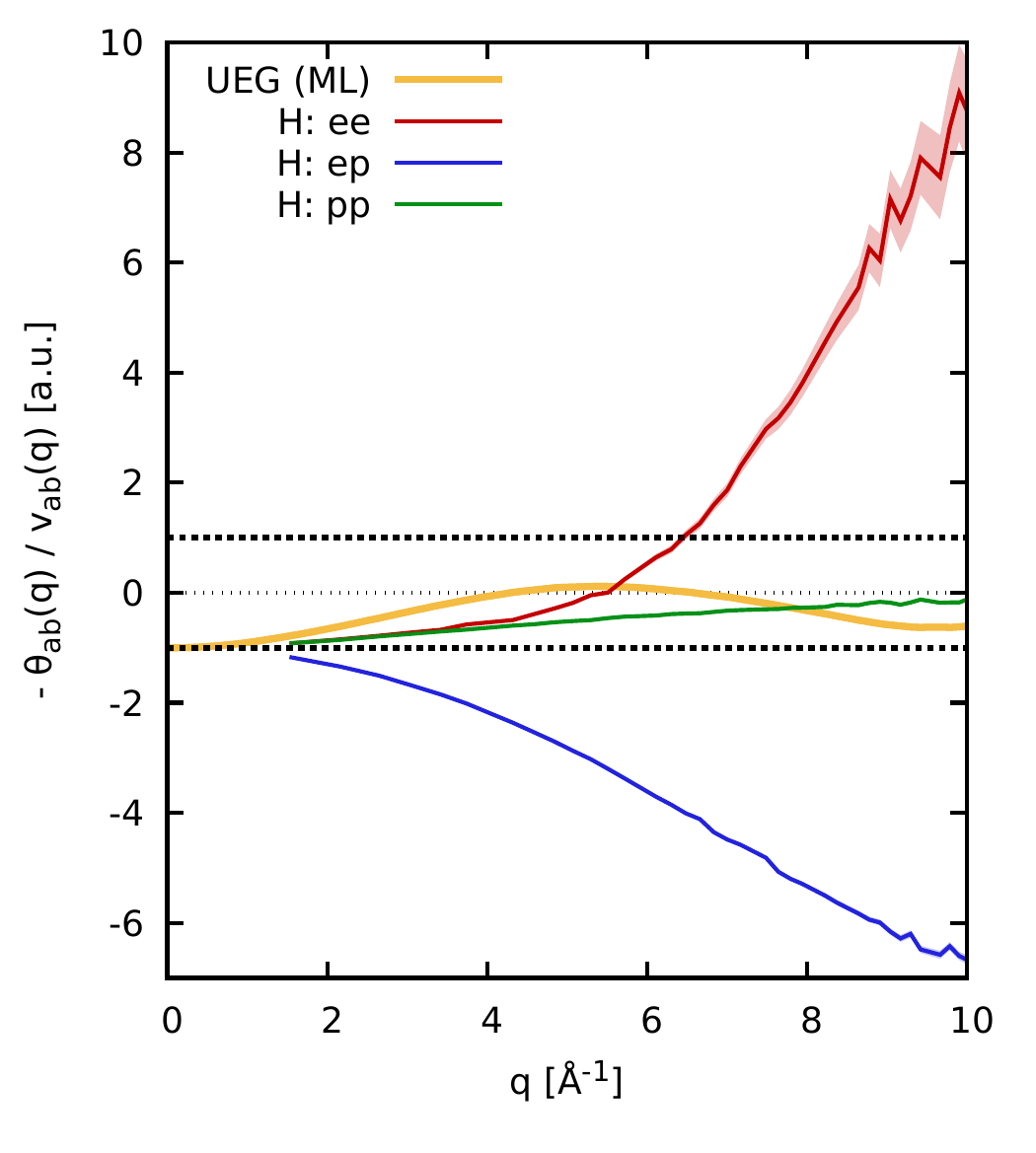}
\caption{\label{fig:LFC_N14_rs2_theta1_q} \emph{Ab initio} PIMC results for the partial local field factors $\theta_{ab}(\mathbf{q})$ for $r_s=2$ and $\Theta=1$. 
Red, green, and blue: $\theta_{ee}(\mathbf{q})$, $\theta_{ep}(\mathbf{q})$, and $\theta_{pp}(\mathbf{q})$ of full hydrogen [Eqs.~(\ref{eq:theta_ee}-\ref{eq:theta_ei})]; yellow: electron--electron local field factor of the UEG model~\cite{dornheim_ML}.
}
\end{figure} 

\begin{figure*}\centering
\hspace*{-0.046\textwidth}\includegraphics[width=0.364\textwidth]{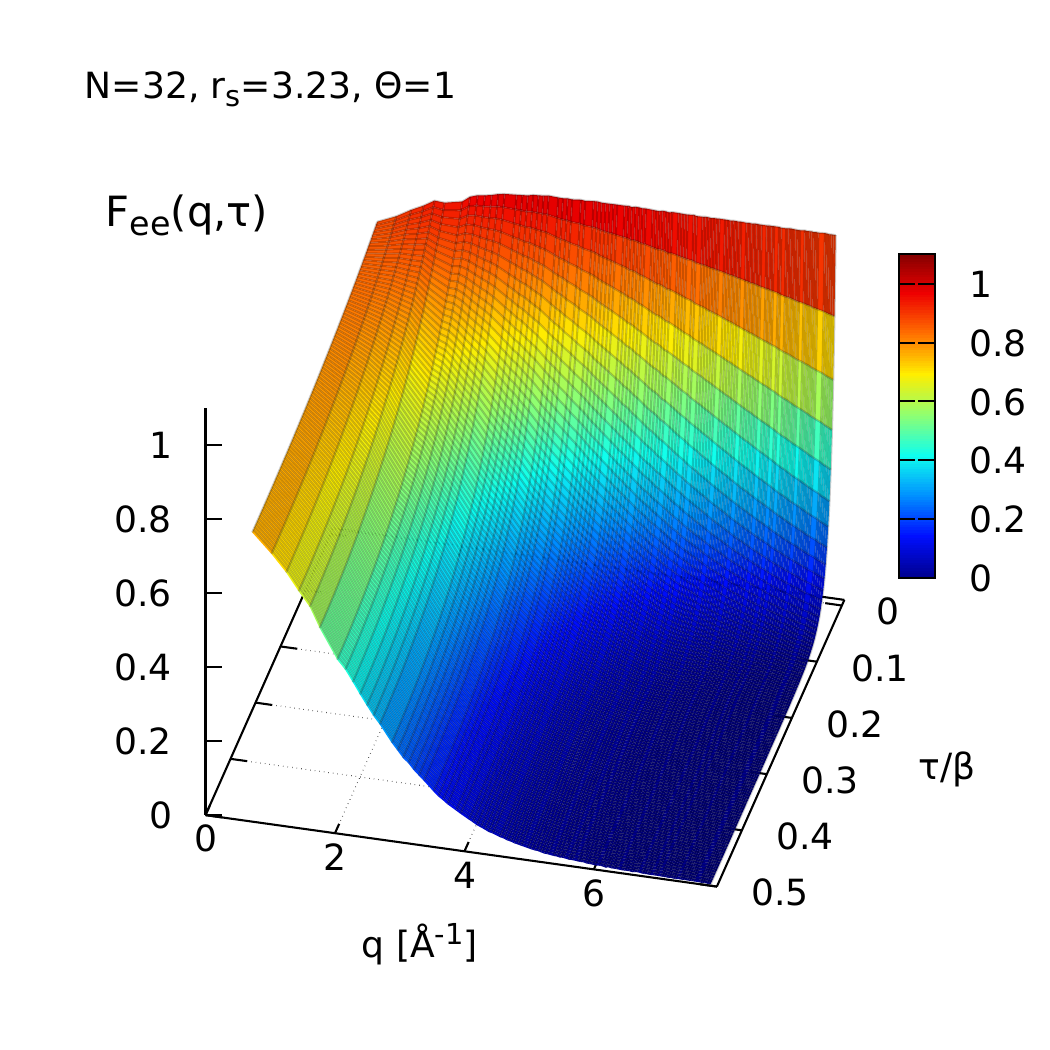}\hspace*{-0.0136\textwidth}\includegraphics[width=0.364\textwidth]{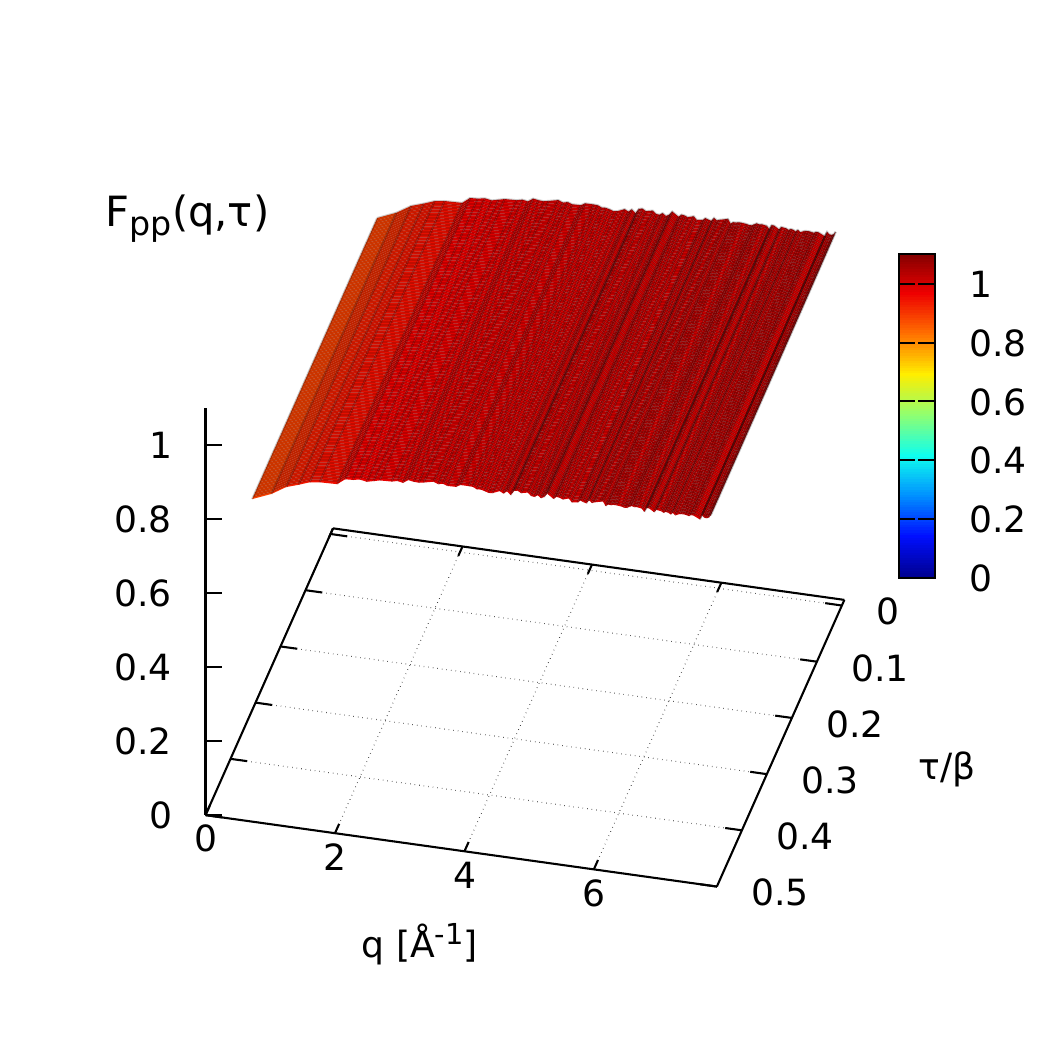}\hspace*{-0.0136\textwidth}\includegraphics[width=0.364\textwidth]{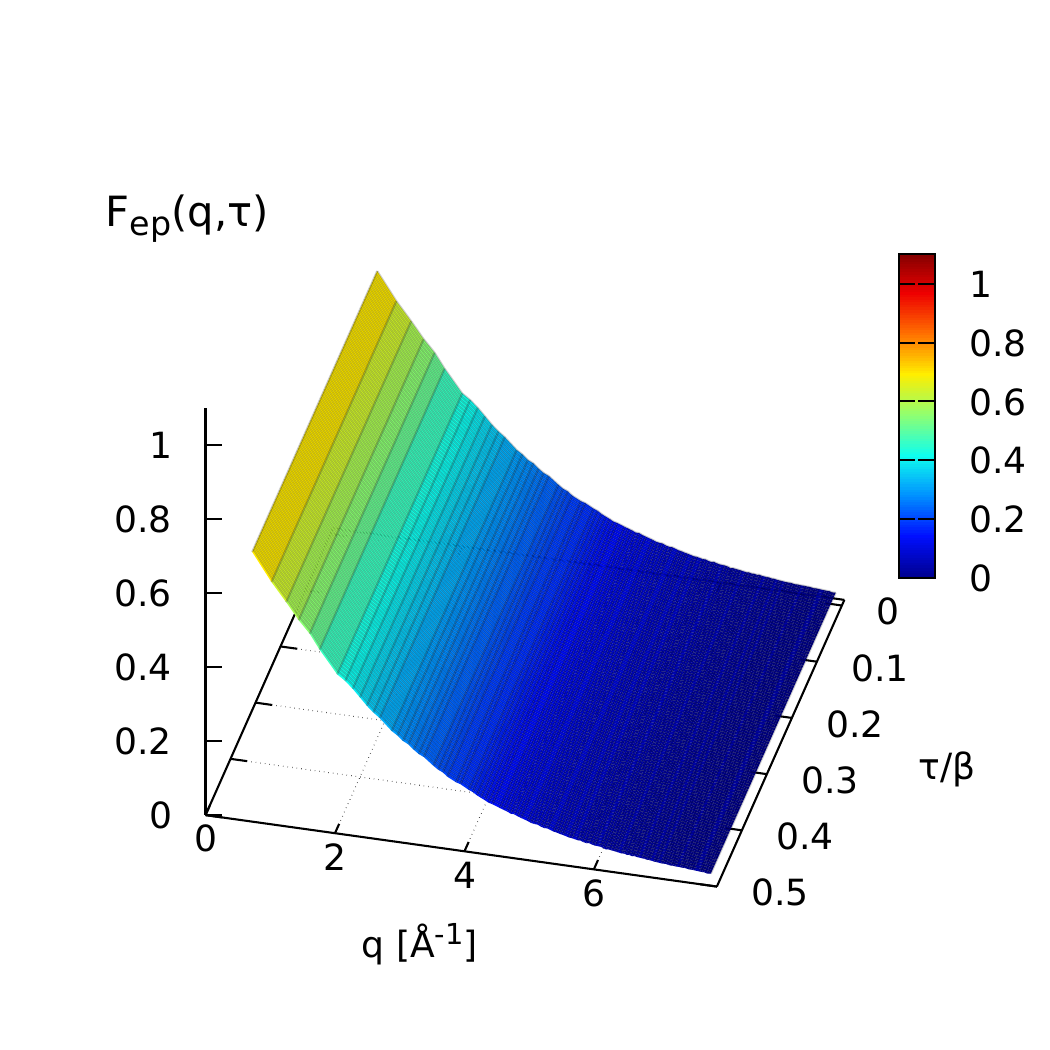}
\caption{\label{fig:ITCF_H_N14_rs3p23_theta1} \emph{Ab initio} PIMC results for the partial imaginary-time density--density correlation functions of warm dense hydrogen for $N=32$ hydrogen atoms at the electronic Fermi temperature $\Theta=1$ and a solid density $r_s=3.23$ in the $\tau$-$q$-plane: electron--electron ITCF $F_{ee}(\mathbf{q},\tau)$ [left], proton--proton ITCF $F_{pp}(\mathbf{q},\tau)$ [center] and electron--proton ITCF $F_{ep}(\mathbf{q},\tau)$ [right].
}
\end{figure*} 

Let us next consider the solid yellow curve in Fig.~\ref{fig:H_N14_rs2_theta1_q}, which shows the density response of the UEG model~\cite{dornheim_ML} at the same conditions.
We find good (though not perfect) agreement with the electronic density response of full hydrogen for $q\gtrsim4\,$\AA$^{-1}$. In stark contrast, there appear substantial differences between the two data sets for $\chi_{ee}(\mathbf{q})$ in the limit of small $q$, where the density response of the UEG vanishes due to perfect screening~\cite{kugler_bounds}. This is not the case for hydrogen due to the electron--proton coupling, which can be understood intuitively in two different ways. From the perspective of density response theory, the fact that we study the response in the static limit of $\omega\to0$ implies that the protons have sufficient time to follow the perturbed electrons, thereby effectively screening the Coulomb interaction between the latter. Ionic mobility breaks down the perfect screening relation of the UEG, and allows the electrons to react even to perturbations on very large length scales (i.e., $q\to0$), directly leading to the nonvanishing value of $\chi_{ee}(\mathbf{q})$ for large wavelengths. Additional insight comes from the relation of $\chi_{ee}(\mathbf{q})$ as the inverse frequency moment of the dynamic structure factor $S_{ee}(\mathbf{q},\omega)$, cf.~Eq.~(\ref{eq:inverse_moment}) in Sec.~\ref{sec:XRTS} above. For the UEG, $S_{ee}(\mathbf{q},\omega)$ simply consists of a sharp plasmon excitation around the (finite) plasma frequency for small $q$, and the weight of the plasmon vanishes quadratically in this regime~\cite{quantum_theory}. For full two-component hydrogen, on the other hand, $S_{ee}(\mathbf{q},\omega)$ contains additional contributions i) from bound-free transitions and ii) from the quasi-elastic feature that is usually modelled 
as a sum of an atomic form factor and a screening cloud of free electrons~\cite{Vorberger_PRE_2015}. The latter feature increases with small $q$ and, being located at very small frequencies, strongly manifests in the inverse moment of $S_{ee}(\mathbf{q},\omega)$; the static electron--electron density response function is thus highly sensitive to electronic localization around the protons. We note that this has potentially important implications for the interpretation of XRTS experiments with WDM, since $\chi_{ee}(\mathbf{q})$ can be directly inferred from the measured intensity (cf.~Sec.~\ref{sec:XRTS}) and, from a theoretical perspective, it does not require dynamic simulations. We thus suggest that, after having inferred the temperature from the model-free ITCF thermometry method introduced in Refs.~\cite{Dornheim_T_2022,Dornheim_T2_2022}, one might calculate $\chi_{ee}(\mathbf{q})$ for a given wavenumber over a relevant interval of densities to infer the latter from the XRTS measurement. Such a strategy would completely circumvent the unphysical decomposition into bound and free electrons, while at the same time being very sensitive to a related, but well-defined concept: electronic localization around the ions. For light elements such as H or Be~\cite{Dornheim_Science_2024}, this might even be accomplished on the basis of quasi-exact PIMC results, offering a pathway for, in principle, approximation-free WDM diagnostics. At the same time, we point out that the static density response might also be estimated with reasonable accuracy from computationally less demanding methods such as DFT or restricted PIMC (using the \emph{direct perturbation approach}) since no dynamic information is required. A dedicated exploration of this idea thus constitutes an important route for future research. Let us conclude our comparison between the electronic density response of full two-component hydrogen and the UEG by inspecting the corresponding density profile in Fig.~\ref{fig:strip_rs2_theta1}, which is shown as the solid yellow curve. As it is expected, the UEG reacts less strongly to an equal external perturbation. Note that the nearly perfect agreement with the green curve is purely coincidental and is a consequence of the intersection of the yellow and blue data in Fig.~\ref{fig:ITCF_H_N14_rs3p23_theta1} for the considered wavenumber.

Finally, the solid black dots in Fig.~\ref{fig:H_N14_rs2_theta1_q} have been adopted from B\"ohme \emph{et al.}~\cite{Bohme_PRL_2022} and show $\chi_{ee}(\mathbf{q})$ of a non-uniform electron gas in the external potential of a fixed proton configuration. Evidently, keeping the protons fixed has a dramatic impact on the electronic density response. Those electrons that are located around a proton are substantially less likely to react to the external harmonic perturbation than the electrons in a free electron gas~\cite{Dornheim_PRE_2023}, leading to an overall reduction of the density response for small to intermediate $q$. In particular, this snapshot based separation of the electrons and protons completely misses the correct signal of the electronic localization around the protons that has been discussed in the previous paragraph. While being ideally suited to benchmark DFT simulations, and potentially to provide input to the latter, the physics content of these results is thus quite incomplete.

Let us conclude our analysis of warm dense hydrogen at $r_s=2$ by considering the various local field factors $\theta_{ab}(\mathbf{q})$, cf.~Eqs.~(\ref{eq:theta_ee})-(\ref{eq:theta_ei}), that are shown in Fig.~\ref{fig:LFC_N14_rs2_theta1_q}. The solid yellow line corresponds to the machine learning (ML) representation of the UEG~\cite{dornheim_ML} and has been included as a reference. The solid red line shows the electron--electron local field factor $\theta_{ee}(\mathbf{q})$. It attains the same limit of $-1$ for $q\to0$, but substantially deviates from the UEG result for all finite wavenumbers. The straightforward application of UEG models for the description of real hydrogen is thus questionable in this regime. This discrepancy becomes even more pronounced for solid state density, but vanishes for $r_s=1$, cf.~Secs.~\ref{sec:solid_density} and \ref{sec:compressed} below.
The green curve shows the proton--proton local field factor $\theta_{pp}(\mathbf{q})$. Interestingly, it is basically indistinguishable from $\theta_{ee}(\mathbf{q})$ for $q\lesssim4\,$\AA$^{-1}$, but diverges from the latter for large wavenumbers. 
Finally, the blue curve shows the electron--proton local field factor $\theta_{ep}(\mathbf{q})$. Unsurprisingly, it is larger in magnitude than the other data sets for small to moderate wavenumbers, which reflects the importance of electron--proton coupling effects.

\subsection{Solid density hydrogen\label{sec:solid_density}}

As a second example, we investigate hydrogen at $\Theta=1$ and $r_s=3.23$, i.e., the density of solid state hydrogen. From a physical perspective, the low density is expected to lead to an increased impact of both electron--electron and electron--proton coupling effects~\cite{low_density1,low_density2,Bohme_PRL_2022}, making these conditions a very challenging test case for simulations. We note that, in contrast to the UEG, the lower density is also more challenging for our PIMC setup. This is a consequence of the incipient formation of H$^{-}$ ions and molecules~\cite{Dornheim_HBe_2024}, leading to a more substantial degree of quantum degeneracy and, consequently, a more severe sign problem. Indeed, we find an average sign of $S\approx0.05$ for $N=14$ hydrogen atoms, causing a factor of $1/S^2=400$ in the required compute time. In addition, low-density hydrogen is expected to exhibit interesting physical effects, such as the \emph{roton-type feature}~\cite{Dornheim_Nature_2022} in the dynamic structure factor $S_{ee}(\mathbf{q},\omega)$ for intermediate wavenumbers~\cite{Hamann_PRR_2023}. Intriguingly, such conditions can be probed with XRTS measurements of optically pumped hydrogen jets~\cite{Zastrau,Fletcher_Frontiers_2022}, which makes our results directly relevant for upcoming experiments.

In Fig.~\ref{fig:ITCF_H_N14_rs3p23_theta1}, we show our new PIMC results for the different ITCFs $F_{ab}(\mathbf{q},\tau)$ in the relevant $\tau$-$q$-plane. Overall, we find the same qualitative trends as with $r_s=2$ investigated above, but with two main differences: i) the electron--proton ITCF $F_{ep}(\mathbf{q},\tau)$ attains larger values on the depicted $q$-grid, indicating a higher degree of coupling between the two species. ii) both $F_{ee}(\mathbf{q},0)$ and $F_{pp}(\mathbf{q},0)$ exhibit a reduced decay for small $q$, for the same reason.

\begin{figure}\centering
\includegraphics[width=0.48\textwidth]{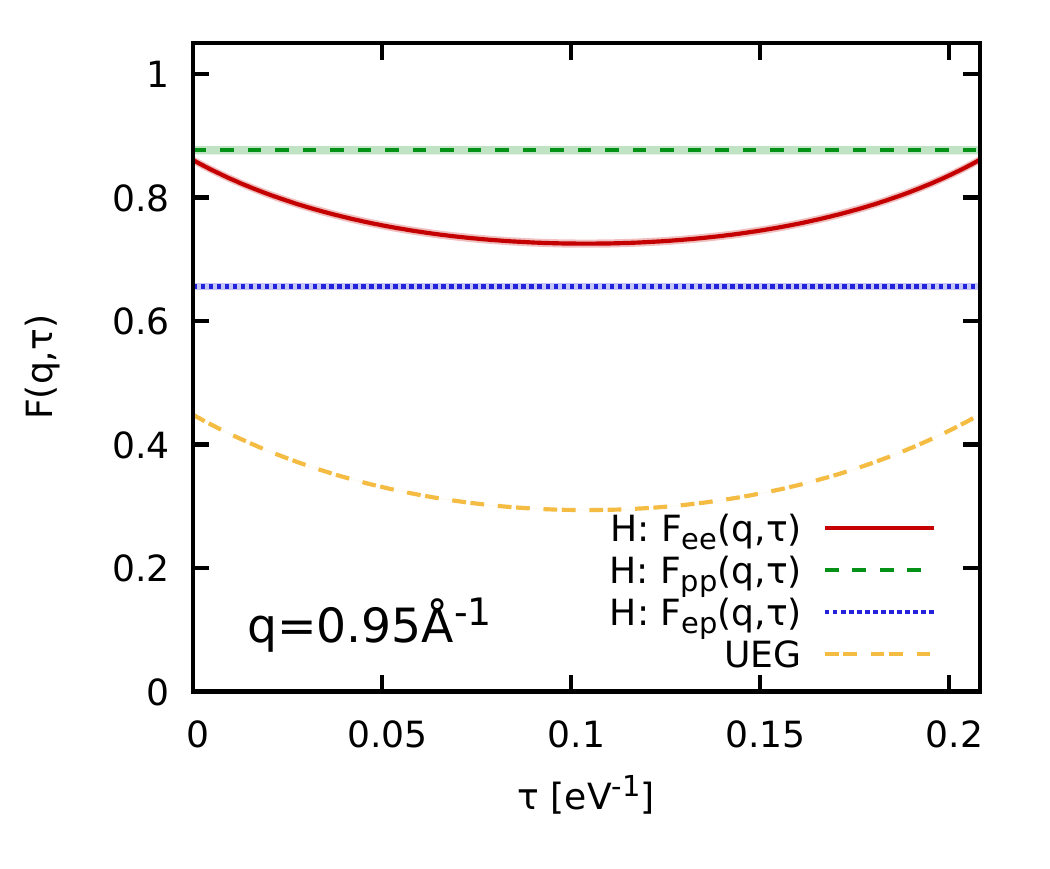}\\\vspace*{-1.3cm}\includegraphics[width=0.48\textwidth]{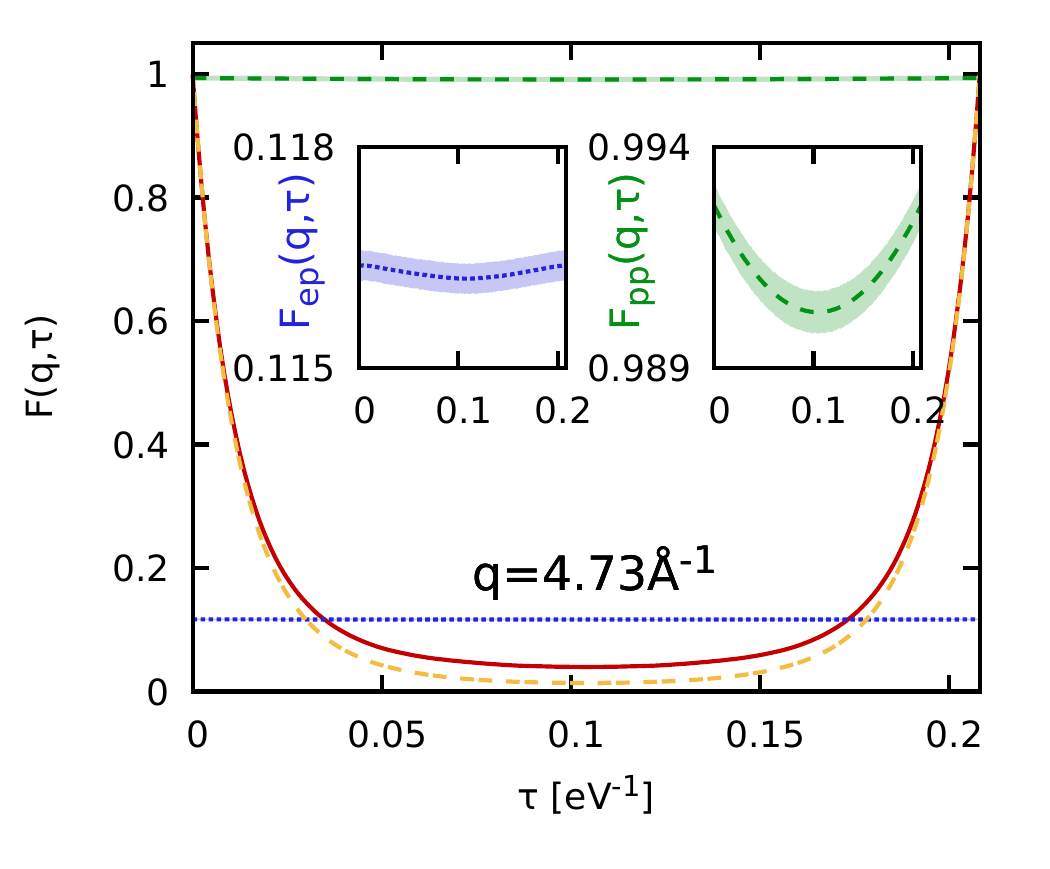}
\caption{\label{fig:H_N14_rs3p23_theta1_ITCF}  \emph{Ab initio} PIMC results for partial hydrogen ITCFs at $r_s=3.23$ and $\Theta=1$ for $q=0.95\,$\AA$^{-1}$ (or $q=0.84q_\textnormal{F}$) [top] and $q=4.73\,$\AA$^{-1}$ (or $q=4.21q_\textnormal{F}$) [bottom]; solid red: $F_{ee}(\mathbf{q},\tau)$, dashed green: $F_{pp}(\mathbf{q},\tau)$, dotted blue: $F_{ep}(\mathbf{q},\tau)$, double-dashed yellow: UEG model~\cite{Dornheim_MRE_2023}. The shaded intervals correspond to $1\sigma$ error bars. The insets in the right panel show magnified segments around $F_{ep}(\mathbf{q},\tau)$ and $F_{pp}(\mathbf{q},\tau)$.
}
\end{figure} 

\begin{figure}\centering
\includegraphics[width=0.47\textwidth]{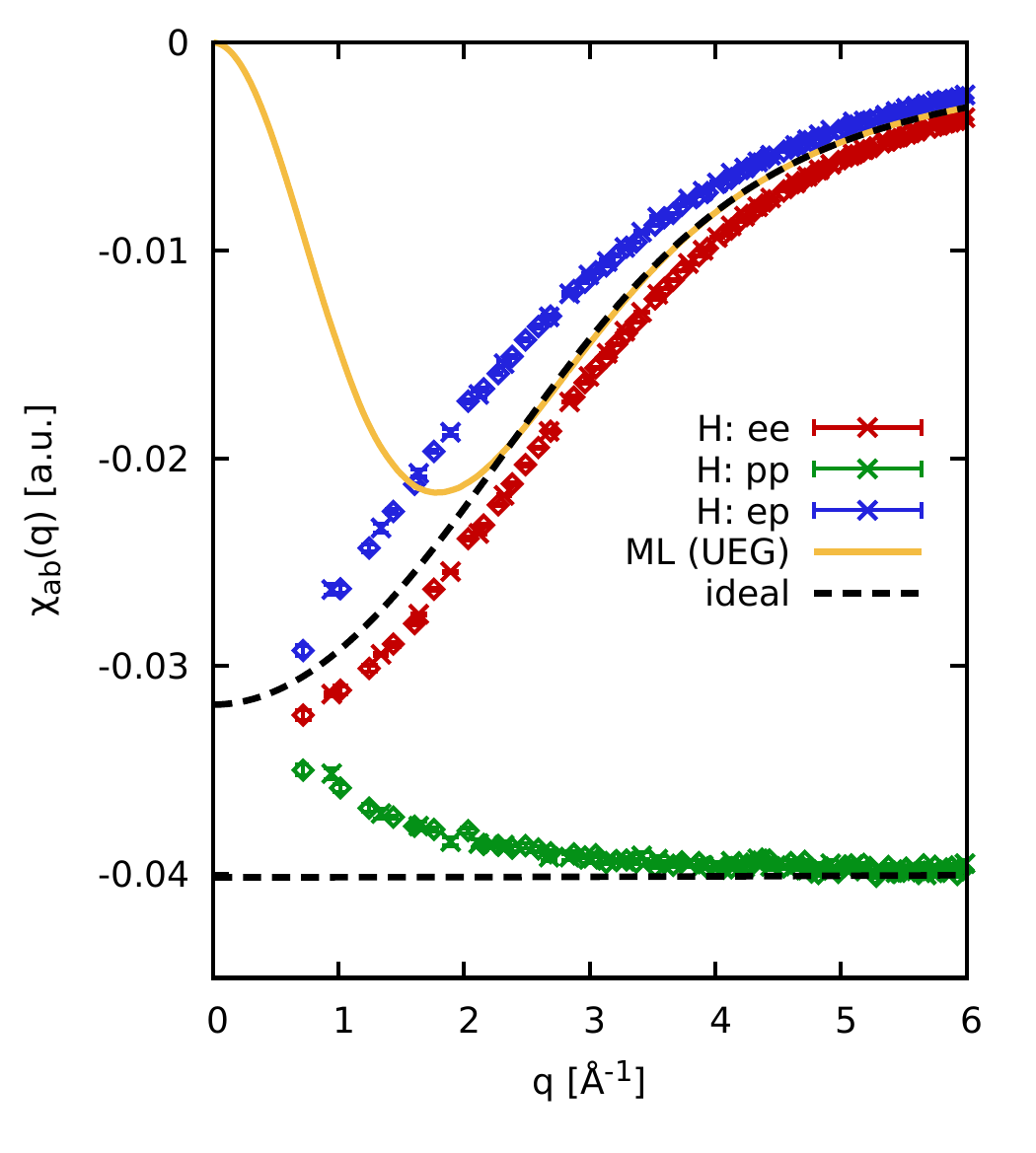}
\caption{\label{fig:H_N14_rs3p23_theta1_q} \emph{Ab initio} PIMC results for the partial static density responses of hydrogen at $r_s=3.23$ and $\Theta=1$. Red, green, blue symbols: $\chi_{ee}(\mathbf{q})$, $\chi_{pp}(\mathbf{q})$, $\chi_{ep}(\mathbf{q})$ for full hydrogen evaluated from the ITCF via Eq.~(\ref{eq:static_chi}); solid yellow line: UEG model~\cite{dornheim_ML}; dashed black: ideal density responses $\chi^{(0)}_{ee}(\mathbf{q})$ and $\chi^{(0)}_{pp}(\mathbf{q})$.
}
\end{figure} 

\begin{figure}\centering\includegraphics[width=0.47\textwidth]{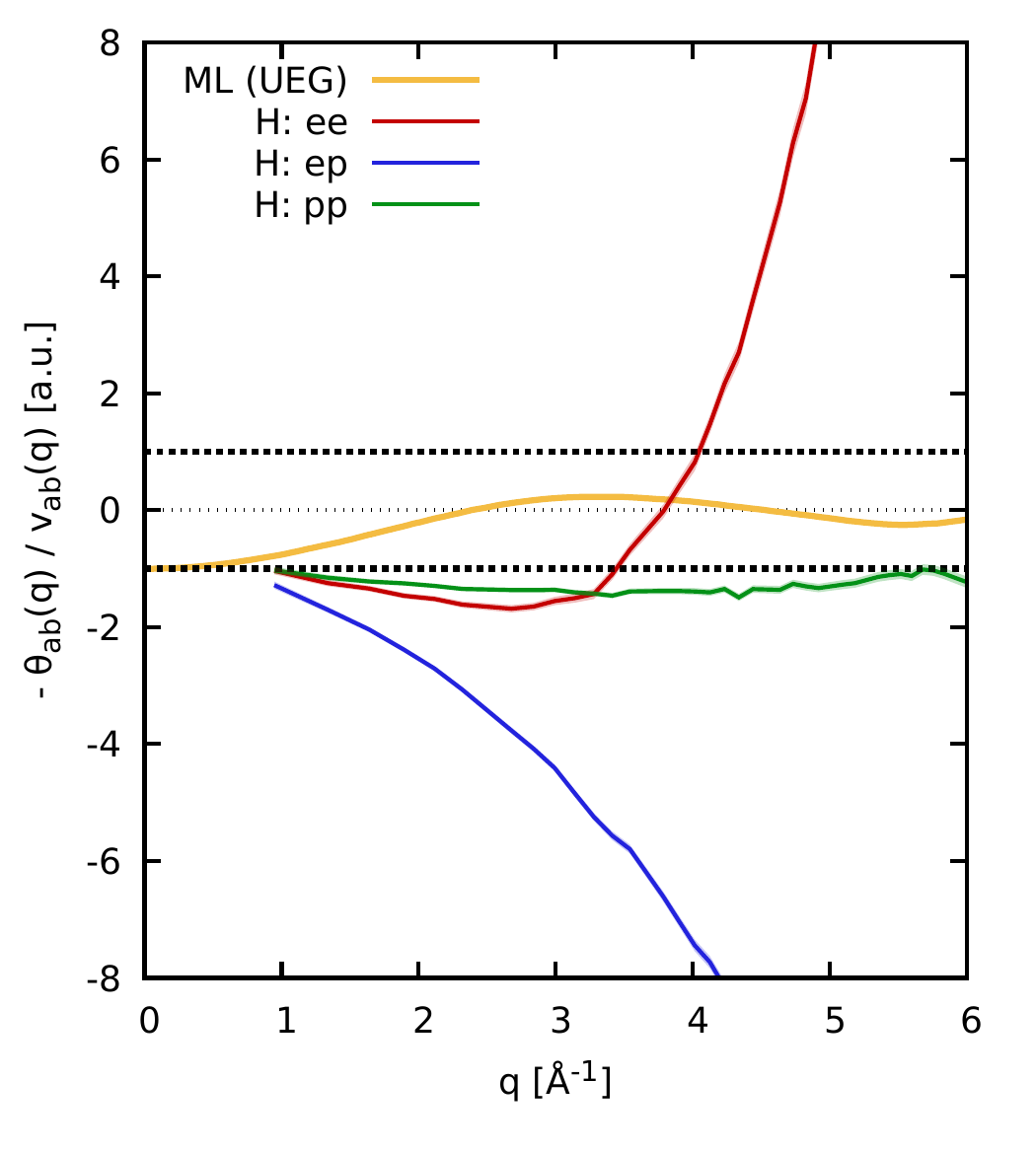}
\caption{\label{fig:LFC_rs3p23} \emph{Ab initio} PIMC results for the partial local field factors $\theta_{ab}(\mathbf{q})$ for $r_s=3.23$ and $\Theta=1$. 
Red, green, and blue: $\theta_{ee}(\mathbf{q})$, $\theta_{ep}(\mathbf{q})$, and $\theta_{pp}(\mathbf{q})$ of full hydrogen [Eqs.~(\ref{eq:theta_ee}-\ref{eq:theta_ei})]; yellow: local field factor of the UEG model~\cite{dornheim_ML}.
}
\end{figure} 

Additional insight comes from Fig.~\ref{fig:H_N14_rs3p23_theta1_ITCF}, where we show the various ITCFs for $q=0.95\,$\AA$^{-1}$ (top) and $q=4.73\,$\AA$^{-1}$ (bottom) along the $\tau$-axis. In the top panel, the main difference from the $r_s=2$ case is the larger offset between the results for $F_{ee}(\mathbf{q},\tau)$ from the UEG and full two-component hydrogen; it is a direct consequence of the increased electronic localization around the protons and the correspondingly increased Rayleigh weight $W_R(\mathbf{q})$.
For the larger $q$-value, we again find that no dependence of $F_{ep}(\mathbf{q},\tau)$ can be resolved within the given confidence interval (shaded blue area). In contrast, we can clearly resolve protonic quantum effects, see the right inset. An additional interesting observation comes from a comparison of the solid red and double-dashed yellow curves corresponding to $F_{ee}(\mathbf{q},\tau)$ for hydrogen and the UEG. In the limit of $\tau\to0$, the two curves are in perfect agreement; this is a consequence of the fact that $S_{ee}(\mathbf{q})\approx1$ and the f-sum rule [Eq.~(\ref{eq:fsum})] yielding the same slope for both data sets. For larger $\tau$, on the other hand, we observe substantial differences between hydrogen and the UEG. Specifically, we find a reduced $\tau$-decay for the former system compared to the latter, which cannot simply be explained by a constant shift due to the Rayleigh weight. From the perspective of our PIMC simulations, this clearly indicates that electron--electron correlations are stabilized along the imaginary-time diffusion process by the presence of the protons. Equivalently, we can attribute this finding to a shift of spectral weight in $S_\textnormal{inel}(\mathbf{q},\omega)$ to lower frequencies (see also Ref.~\cite{Dornheim_MRE_2023} for a more detailed discussion), indicating a nontrivial structure of the full DSF. The presented data for $F_{ee}(\mathbf{q},\tau)$ thus constitute rigorous benchmarks for models (e.g.,~the Chihara decomposition) and simulations (e.g.,~LR-TDDFT), and a dedicated future comparative analysis will give important insights into the validity range of different methods.

\begin{figure*}\centering
\hspace*{-0.046\textwidth}\includegraphics[width=0.364\textwidth]{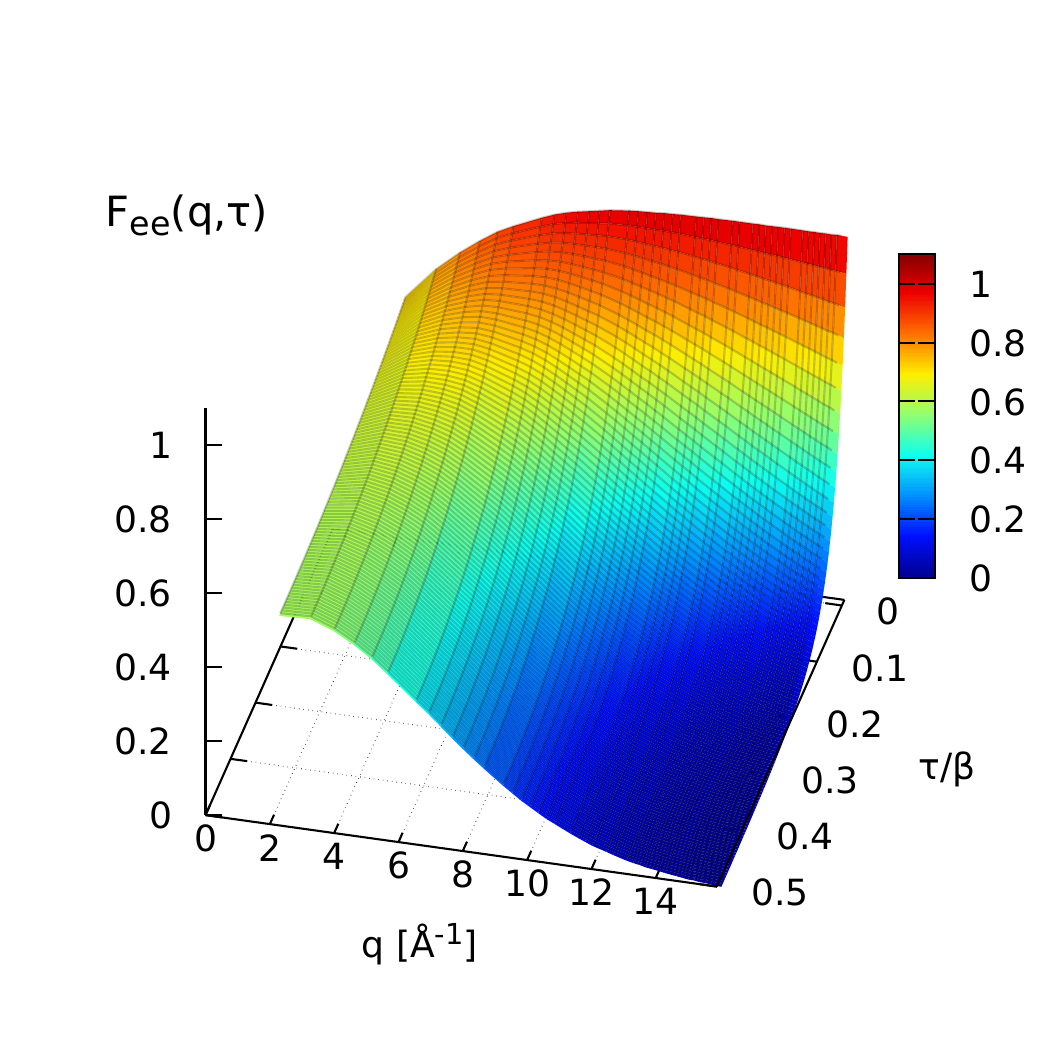}\hspace*{-0.0136\textwidth}\includegraphics[width=0.364\textwidth]{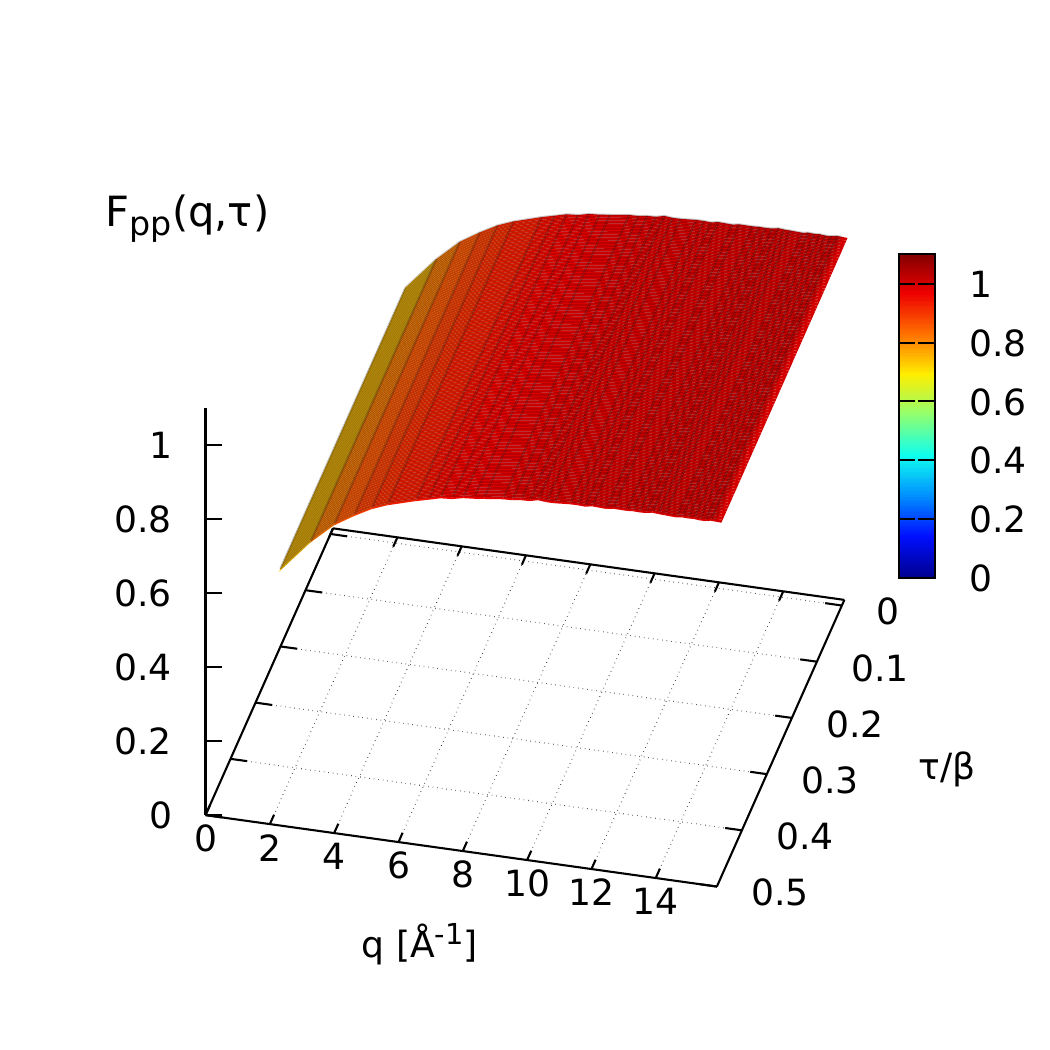}\hspace*{-0.0136\textwidth}\includegraphics[width=0.364\textwidth]{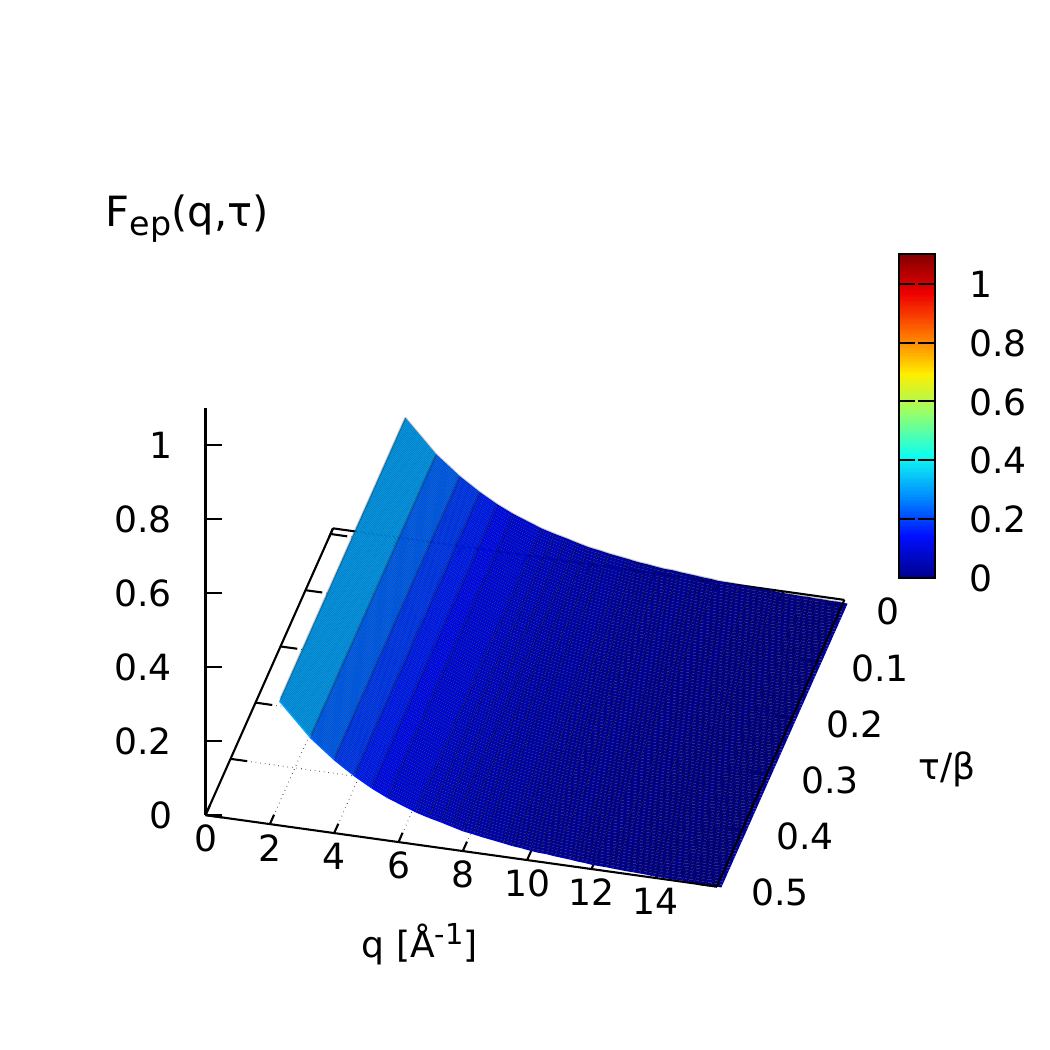}
\caption{\label{fig:ITCF_H_N14_rs1_theta1} \emph{Ab initio} PIMC results for the partial imaginary-time density--density correlation functions of warm dense hydrogen for $N=32$ hydrogen atoms at the electronic Fermi temperature $\Theta=1$ and a compressed density $r_s=1$ in the $\tau$-$q$-plane: electron--electron ITCF $F_{ee}(\mathbf{q},\tau)$ [left], proton--proton ITCF $F_{pp}(\mathbf{q},\tau)$ [center] and electron--proton ITCF $F_{ep}(\mathbf{q},\tau)$ [right].
}
\end{figure*}

In Fig.~\ref{fig:H_N14_rs3p23_theta1_q}, we show the corresponding species-resolved static density response functions $\chi_{ab}(\mathbf{q})$, and again restrict ourselves to a discussion of the main differences from the $r_s=2$ case. i) $\chi_{ee}(\mathbf{q})$ monotonically increases with decreasing $q$ in over the entire depicted $q$-range, and the electronic localization around the protons predominantly shapes its behaviour for small $q$. ii) in contrast to the UEG, $\chi_{ee}(\mathbf{q})$ does not converge towards the ideal density response $\chi^{(0)}_{ee}(\mathbf{q})$ for large $q$, which is a direct consequence of the reduced $\tau$-decay of $F_{ee}(\mathbf{q},\tau)$ depicted in the bottom panel of Fig.~\ref{fig:ITCF_H_N14_rs3p23_theta1} above. iii) $\chi_{ep}(\mathbf{q})$ is substantially larger, with respect to $\chi_{ee}(\mathbf{q})$ and $\chi_{pp}(\mathbf{q})$, for $r_s=3.23$ compared to $r_s=2$. iv) $\chi_{pp}(\mathbf{q})$ exhibits a reduced decay for small $q$ compared to $r_s=2$; this is due to the electronic screening of the proton--proton interaction, making the proton response more ideal.

Finally, we show the partial local field factors $\theta_{ab}(\mathbf{q})$ in Fig.~\ref{fig:LFC_rs3p23}. While the electron--electron local field factor of full two-component hydrogen (solid red) attains the same limit as the UEG model~\cite{dornheim_ML} (yellow) for $q\to0$, it exhibits the opposite trend for intermediate $q$, followed by a steep increase for shorter wavelengths. This is consistent with previous findings by B\"ohme \emph{et al.}~\cite{Bohme_PRL_2022} for an electron gas in a fixed external proton potential at similar conditions ($r_s=4$) and clearly indicates the breakdown of UEG models when electrons are strongly localized. The proton--proton local field factor is relatively featureless over the entire $q$-range, which might be due to the aforementioned electronic screening of the proton--proton interaction. Similar to $r_s=2$, $\theta_{ep}(\mathbf{q})$
constitutes the largest local field factor for relevant wave numbers.

\subsection{Strongly compressed hydrogen\label{sec:compressed}}

\begin{figure}\centering
\includegraphics[width=0.48\textwidth]{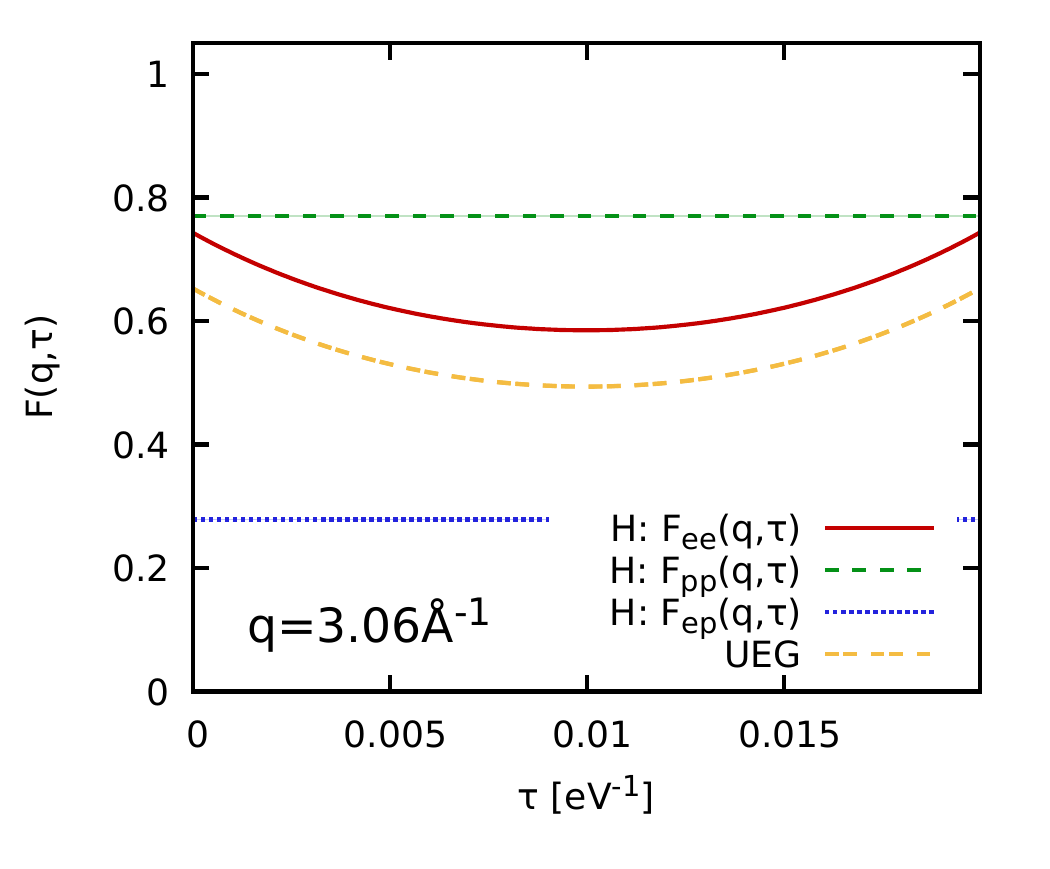}\\\vspace*{-1.3cm}\includegraphics[width=0.48\textwidth]{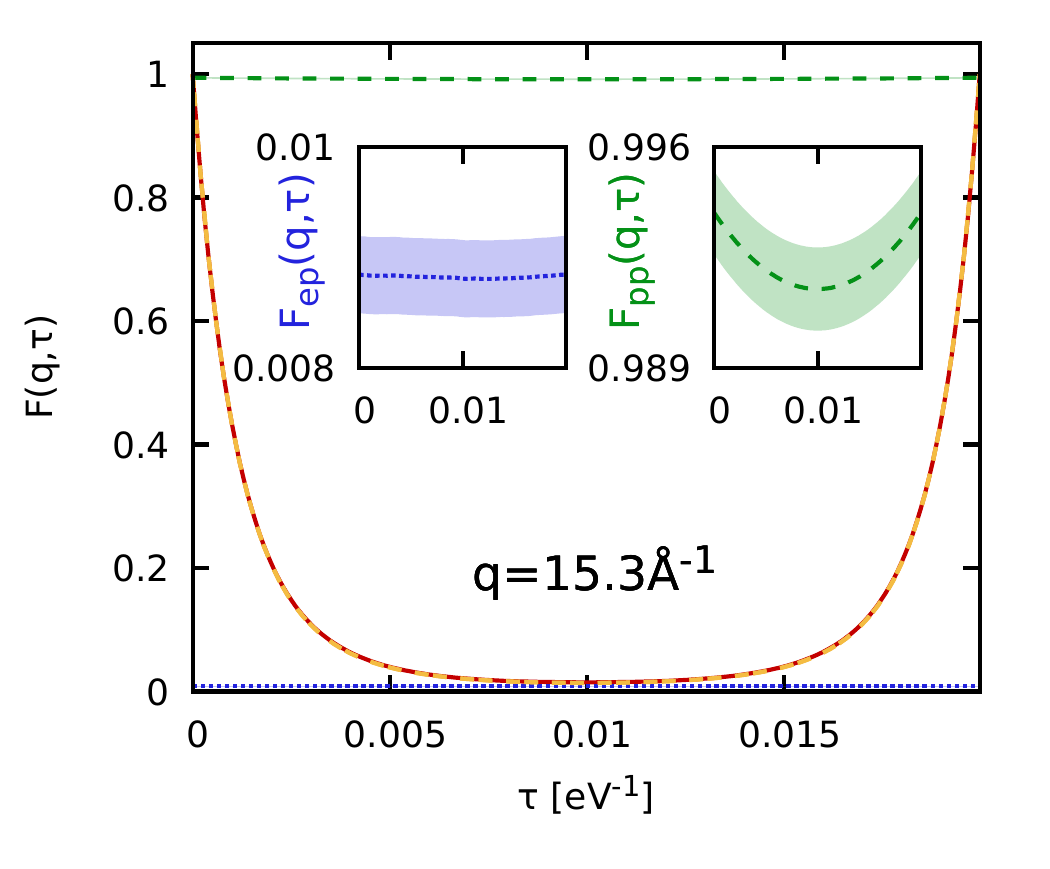}
\caption{\label{fig:H_N14_rs1_theta1_ITCF}  \emph{Ab initio} PIMC results for partial hydrogen ITCFs at $r_s=1$ and $\Theta=1$ for $q=3.06\,$\AA$^{-1}$ (or $q=0.84q_\textnormal{F}$) [top] and $q=15.3\,$\AA$^{-1}$ (or $q=4.21q_\textnormal{F}$) [bottom]; solid red: $F_{ee}(\mathbf{q},\tau)$, dashed green: $F_{pp}(\mathbf{q},\tau)$, dotted blue: $F_{ep}(\mathbf{q},\tau)$, double-dashed yellow: UEG model~\cite{Dornheim_MRE_2023}. The shaded intervals correspond to $1\sigma$ error bars. The insets in the right panel show magnified segments around $F_{ep}(\mathbf{q},\tau)$ and $F_{pp}(\mathbf{q},\tau)$.
}
\end{figure} 

As the final example, we investigate compressed hydrogen at $\Theta=1$ and $r_s=1$. The corresponding PIMC results for the species-resolved ITCFs are shown in Fig.~\ref{fig:ITCF_H_N14_rs1_theta1} in the $\tau$-$q$-plane and qualitatively closely resemble the case of $r_s=2$ shown in Fig.~\ref{fig:H_N14_rs2_theta1_ITCF} above. The main difference is the reduced magnitude of $F_{ep}(\mathbf{q},\tau)$, indicating a substantially weaker localization of the electrons around the protons, as it is expected.
In Fig.~\ref{fig:H_N14_rs1_theta1_ITCF}, we show the ITCFs along the $\tau$-direction for $q=3.06\,$\AA$^{-1}$ and $q=15.3\,$\AA$^{-1}$, i.e, for the same values of $q/q_\textnormal{F}$ as in Figs.~\ref{fig:H_N14_rs2_theta1_ITCF} and \ref{fig:H_N14_rs3p23_theta1_ITCF}. We find very similar behaviour as for $r_s=2$ with a reduced electron--proton coupling. In fact, no deviations can be resolved between $F_{ee}(\mathbf{q},\tau)$ for the UEG model and full two-component hydrogen in the bottom panel. In other words, compressed hydrogen closely resembles a free electron gas.

\begin{figure}\centering
\includegraphics[width=0.48\textwidth]{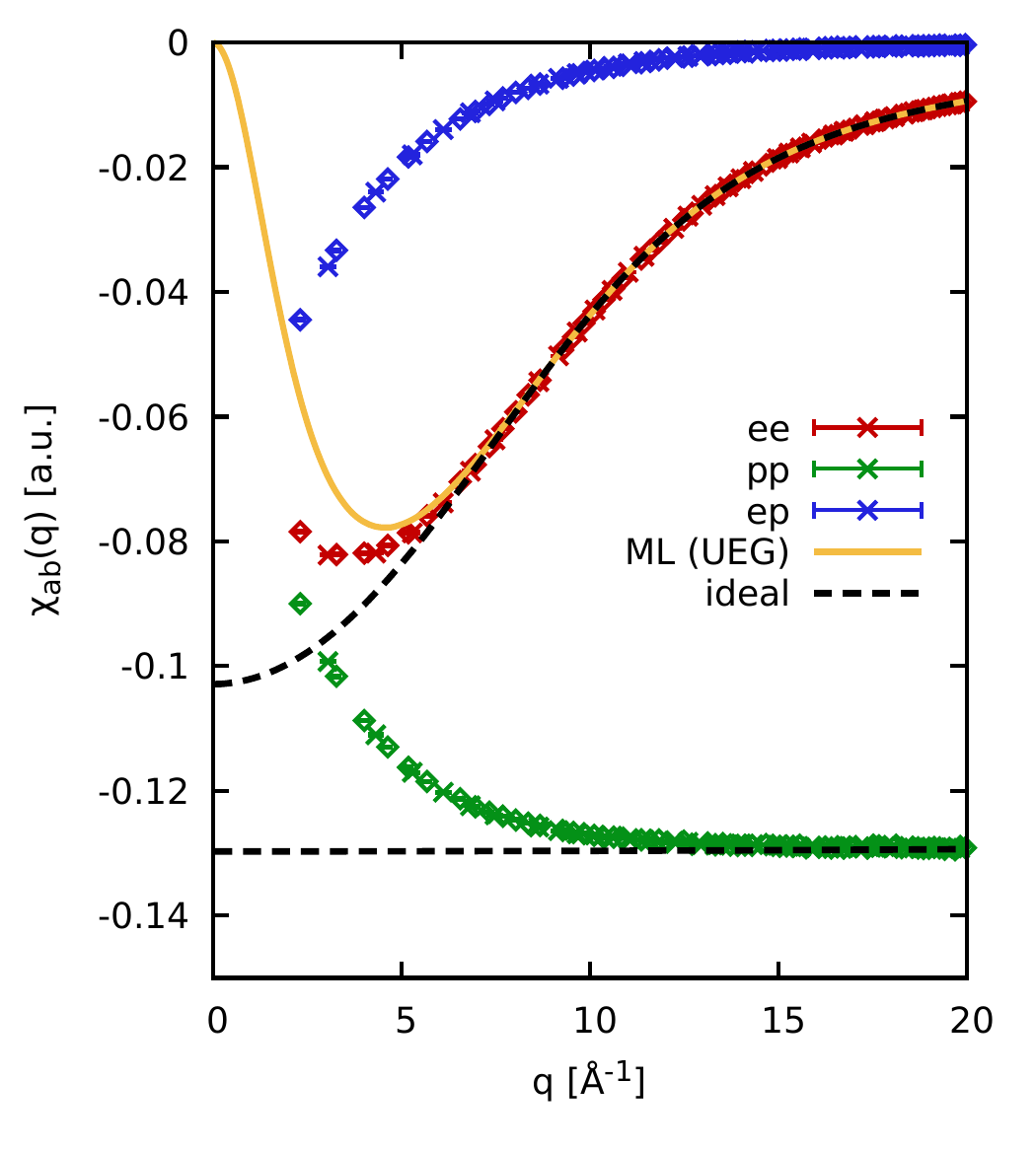}
\caption{\label{fig:H_N14_rs1_theta1_q} \emph{Ab initio} PIMC results for the partial static density responses of hydrogen at $r_s=1$ and $\Theta=1$. Red, green, blue symbols: $\chi_{ee}(\mathbf{q})$, $\chi_{pp}(\mathbf{q})$, $\chi_{ep}(\mathbf{q})$ for full hydrogen evaluated from the ITCF via Eq.~(\ref{eq:static_chi}); solid yellow line: UEG model~\cite{dornheim_ML}; dashed black: ideal density responses $\chi^{(0)}_{ee}(\mathbf{q})$ and $\chi^{(0)}_{pp}(\mathbf{q})$.
}
\end{figure} 

\begin{figure}\centering
\includegraphics[width=0.48\textwidth]{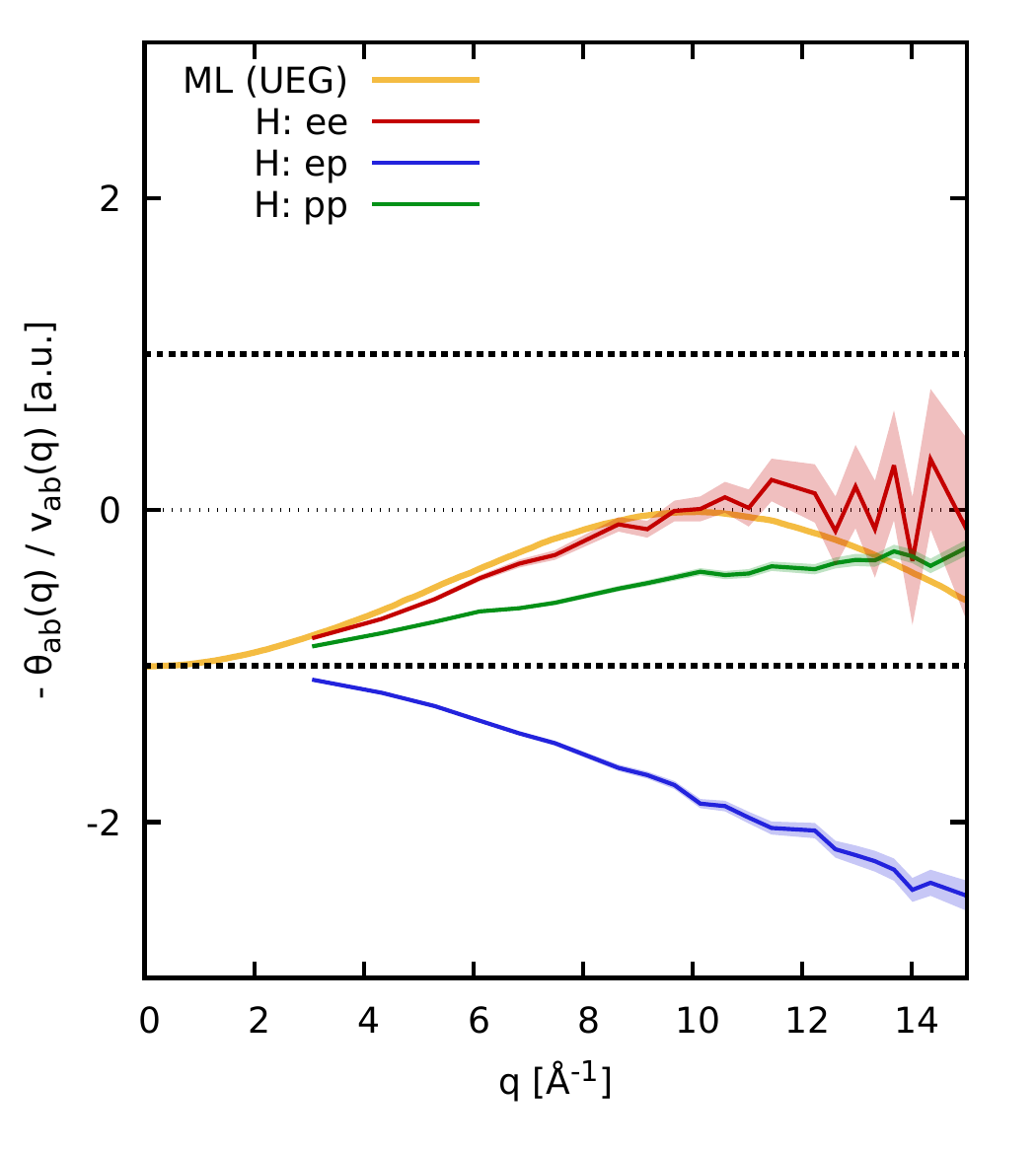}
\caption{\label{fig:H_N14_rs1_theta1_theta} \emph{Ab initio} PIMC results for the partial local field factors $\theta_{ab}(\mathbf{q})$ for $r_s=1$ and $\Theta=1$. 
Red, green, and blue: $\theta_{ee}(\mathbf{q})$, $\theta_{ep}(\mathbf{q})$, and $\theta_{pp}(\mathbf{q})$ of full hydrogen [Eqs.~(\ref{eq:theta_ee}-\ref{eq:theta_ei})]; yellow: local field factor of the UEG model~\cite{dornheim_ML}.
}
\end{figure}

This observation is further substantiated in Fig.~\ref{fig:H_N14_rs1_theta1_q}, where we show the corresponding partial static density response functions $\chi_{ab}(\mathbf{q})$. Evidently, $\chi_{ee}(\mathbf{q})$ is very similar to the UEG, and converges towards the latter for $q\gtrsim5\,$\AA$^{-1}$. Nevertheless, we can still clearly detect the signature of electronic localization around the protons as a somewhat increased response for smaller $q$.  In fact, the convergence of the electronic density response of hydrogen towards the UEG model for high densities can be seen most clearly in Fig.~\ref{fig:H_N14_rs1_theta1_theta}
where we show the partial local field factors $\theta_{ab}(\mathbf{q})$. In stark contrast to $r_s=2$ and mainly to $r_s=3.23$, the UEG model is in very good agreement with the true $\theta_{ee}(\mathbf{q})$ of hydrogen at $r_s=1$. This has important implications for laser fusion applications and clearly indicates that UEG based models such as the adiabatic local density approximation are appropriate over substantial parts of the ICF compression path. Finally, we note that the electron--proton local field factor $\theta_{ep}(\mathbf{q})$ has the same magnitude as $\theta_{ee}(\mathbf{q})$ in this case, whereas $\theta_{pp}(\mathbf{q})$ is somewhat smaller.

\section{Summary and Discussion\label{sec:summary}}

In this work, we have presented the first \emph{ab initio} results for the partial density response functions of warm dense hydrogen. This has been achieved on the basis of direct PIMC simulations that are computationally very expensive owing to the fermion sign problem, but exact within the given Monte Carlo error bars. Moreover, we have employed the recently introduced $\xi$-extrapolation technique~\cite{Xiong_JCP_2022,Dornheim_JCP_2023,Dornheim_JPCL_2024,Dornheim_Science_2024,Dornheim_HBe_2024} to access larger system sizes; no finite-size effects have been detected for the wavenumber-resolved properties, in agreement with previous results for the UEG model at the same conditions~\cite{dornheim_prl}. A particular advantage of the direct PIMC method is that it allows us to estimate all ITCFs $F_{ab}(\mathbf{q},\tau)$. First and foremost, this gives us direct access to the full wavenumber dependence of the static density response $\chi_{ab}(\mathbf{q})$ and, consequently, the local field factor $\theta_{ab}(\mathbf{q})$ from a single simulation of the unperturbed system. As an additional crosscheck, we have also carried out extensive simulations of hydrogen where either the electrons or the protons are subject to an external monochromatic perturbation. We find perfect agreement between the \emph{direct perturbation approach} and the ITCF-based method in the linear response regime, as it is expected.

In addition to the anticipated impact on future investigations of WDM that is outlined below, the presented study is highly interesting in its own right and has given new insights into the complex interplay of the electrons and protons in different regimes. We repeat that both $F_{ee}(\mathbf{q},\tau)$ and $\chi_{ee}(\mathbf{q})$ can be obtained from XRTS measurements, and our results thus constitute unambiguous predictions for experiments with ICF plasmas and hydrogen jets. In particular, we have shown that $\chi_{ee}(\mathbf{q})$ is highly sensitive to the electronic localization around the protons. This effect is particularly pronounced for small wavenumbers, which can be probed in forward scattering geometries~\cite{siegfried_review}, and it is directly related to the important concept of effective ionization. At the same time, we stress that the latter is, strictly speaking, ill-defined and ambiguous, whereas the reported impact of electronic localization on $\chi_{ee}(\mathbf{q})$ constitutes a well-defined physical observable both in experiments and simulations. In terms of physical parameters, we have found that the electrons in hydrogen behave very differently from the UEG model for $r_s=2$ and even more so for $r_s=3.23$, while the UEG model is appropriate for strongly compressed hydrogen at $r_s=1$. Finally, we have reported, to the best of our knowledge, the first reliable results for the local field factor $\theta_{ab}(\mathbf{q})$ that is directly related to the XC-kernel from LR-TDDFT calculations.

We are convinced that our study opens up a wealth of opportunities for impactful future work, and helps to further lift WDM theory onto the level of true predictive capability. In the following, we give a non-exhaustive list of particularly promising projects.

i) The electron--electron density response function $\chi_{ee}(\mathbf{q})$ is ideally suited for the model-free interpretation of XRTS measurements of WDM. Specifically, we propose to first infer the temperature from the exact ITCF-based thermometry method introduced in Refs.~\cite{Dornheim_T_2022,Dornheim_T2_2022} and to subsequently carry out a set of PIMC simulations for $\chi_{ee}(\mathbf{q})$ over a relevant set of densities. Matching the PIMC result to the experimental result for $\chi_{ee}(\mathbf{q})$, the inverse moment of $S_{ee}(\mathbf{q},\omega)$, then gives one model-free access to the density. This PIMC based interpretation framework of XRTS experiments will give new insights into the equation-of-state of WDM, with important implications for astrophysical models and laser fusion applications. Additionally, we note that computing $\chi_{ee}(\mathbf{q})$ does not require dynamical simulations or dynamic XC-kernels, and, therefore, might be suitable for approximate methods such as DFT.

ii) The species-resolved local field factors $\theta_{ab}(\mathbf{q})$, and in particular the electron--electron XC-kernel, constitute key input for a gamut of applications, including the estimation of thermal and electrical conductivities~\cite{Veysman_PRE_2016}, the construction of effective potentials~\cite{ceperley_potential,Kukkonen_PRB_2021,Dornheim_JCP_2022}, and the estimation of the ionization potential depression~\cite{Zan_PRE_2021}. Two particularly enticing applications concern the construction of nonlocal XC-functionals for DFT simulations based on the adiabatic connection formula and the fluctuation--dissipation theorem~\cite{pribram}, and LR-TDDFT simulations within the adiabatic approximation~\cite{Moldabekov_PRR_2023,ullrich2012time}. In fact, previous studies of the UEG~\cite{dornheim_dynamic,dynamic_folgepaper,Dornheim_PRL_2020_ESA,Dornheim_PRB_2021} have reported that the utilization of a static XC-kernel is capable of giving highly accurate results for $S_{ee}(\mathbf{q},\omega)$ over substantial parts of the WDM regime. Extending these considerations for hydrogen and beyond are promising routes that will be explored in dedicated future works.

iii) Our quasi-exact PIMC results constitute a rigorous benchmark for computationally less expensive though approximate simulation methods, most importantly thermal DFT. Therefore, a dedicated comparative investigation for a real WDM system will give invaluable new insights into the range of applicability of available XC-functionals, and guide the development of new thermal functionals that are explicitly designed for application in the WDM regime~\cite{ksdt,groth_prl,Karasiev_PRL_2018,Karasiev_PRB_2022,kozlowski2023generalized}. In addition, the presented results for both $F_{ee}(\mathbf{q},\tau)$ and $\chi_{ee}(\mathbf{q})$ can be used to benchmark dynamic methods such as LR-TDDFT~\cite{Schoerner_PRE_2023,Moldabekov_PRR_2023}, real-time TDDFT~\cite{Baczewski_PRL_2016}, or indeed the popular but uncontrolled Chihara models~\cite{Chihara_1987,Gregori_PRE_2003,siegfried_review,boehme2023evidence}.

iv) A less straightforward, though highly rewarding endeavour is given by the so-called \emph{analytic continuation} of $F_{ee}(\mathbf{q},\tau)$, i.e., the numerical inversion of Eq.~(\ref{eq:Laplace}) to solve for the dynamic structure factor $S_{ee}(\mathbf{q},\omega)$. While such an inverse Laplace transform constitutes a notoriously difficult and, in fact, ill-posed problem~\cite{JARRELL1996133}, this issue has been circumvented recently for the warm dense UEG based on the stochastic sampling of the dynamic local field factor with rigorous constraints imposed on the trial solutions~\cite{dornheim_dynamic,dynamic_folgepaper}. Finding similar constraints for warm dense hydrogen would open up the way for the first exact results for the DSF of real WDM systems as well as for related properties such as the dynamic dielectric function, conductivity, and dynamic density response~\cite{Hamann_PRB_2020,Hamann_CPP_2020}.

v) Finally, we note that current direct PIMC capabilities allow for highly accurate simulations of elements up to Be~\cite{Dornheim_Science_2024,Dornheim_HBe_2024}. It will be very interesting to see how the more complex behaviour of such systems (e.g.,~double occupation of the K shell of Be for $T=100\,$eV~\cite{Dornheim_Science_2024,Dornheim_HBe_2024}) manifests in the different ITCFs and density response functions. Moreover, these considerations might be extended to material mixtures such as LiH, giving rise to additional cross correlations that can be straightforwardly estimated by upcoming PIMC simulations.


\appendix
\section{Fermion sign problem and $\xi$-extrapolation method}\label{sec:appendix}

\begin{figure}\centering
\includegraphics[width=0.48\textwidth]{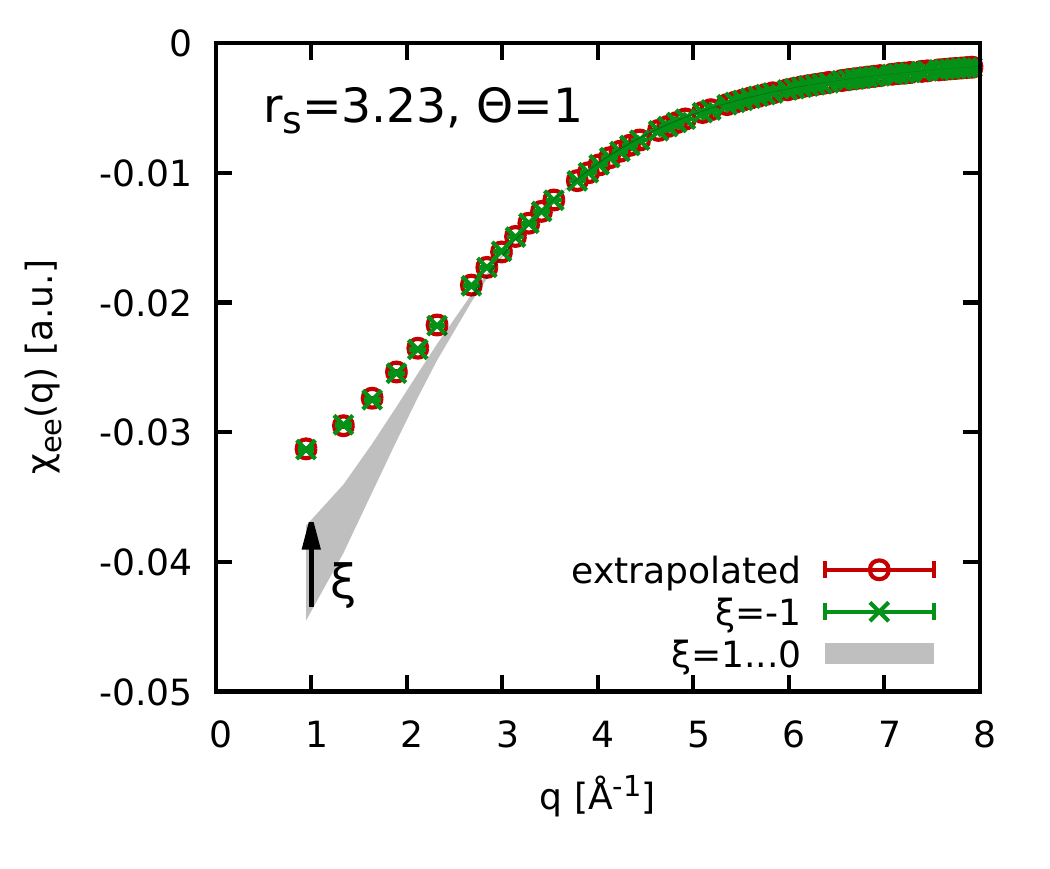}\\\vspace*{-1cm}
\includegraphics[width=0.48\textwidth]{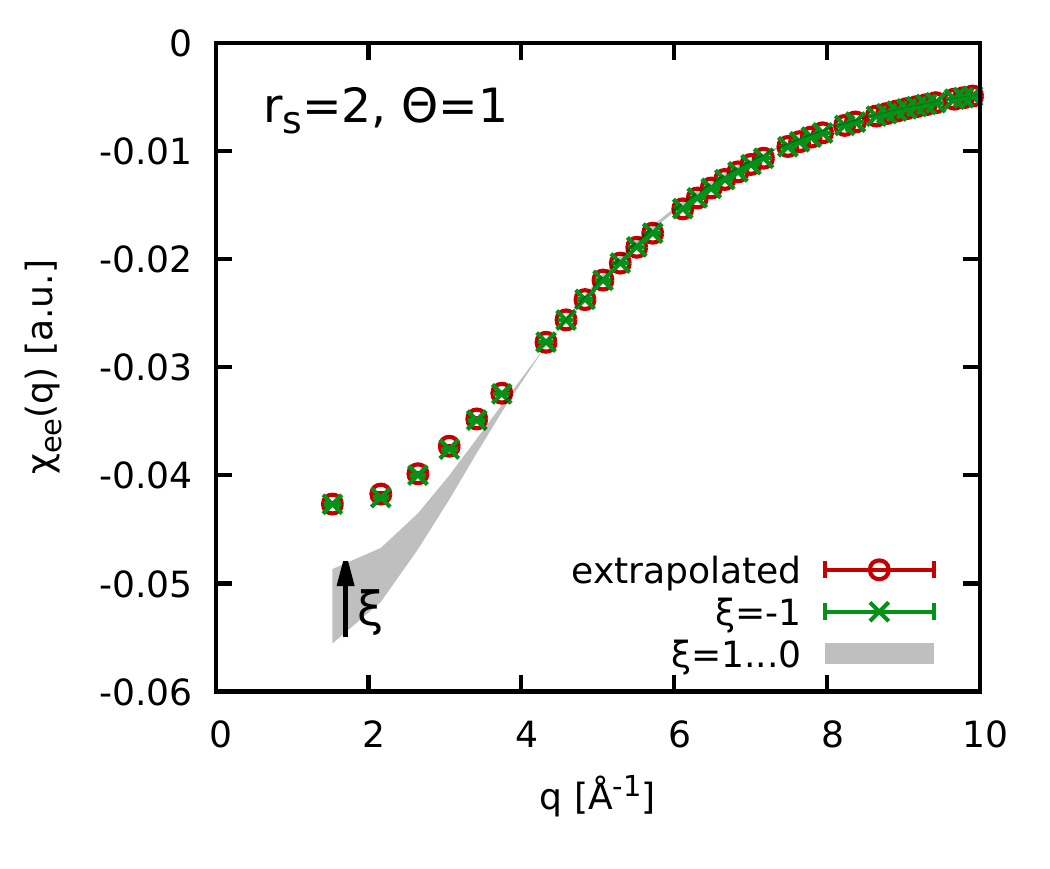}\\\vspace*{-1cm}
\includegraphics[width=0.48\textwidth]{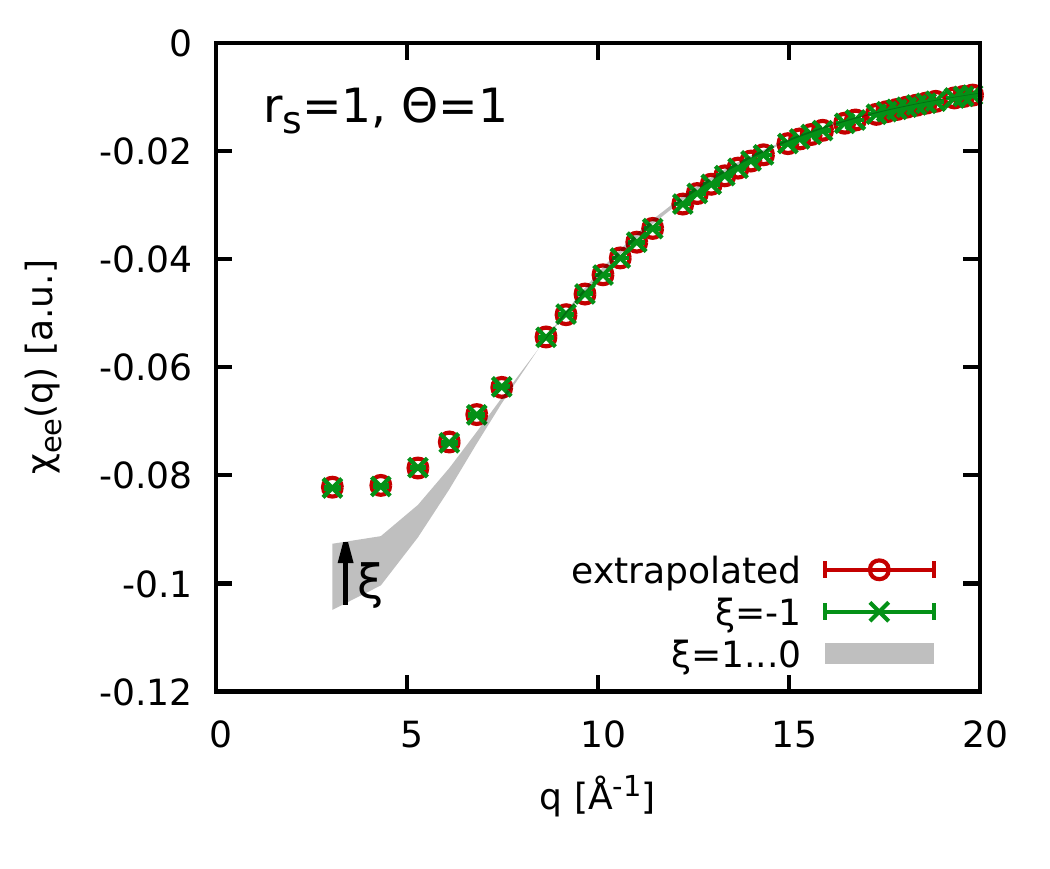}\\
\caption{\label{fig:H_N14_rs2_theta1_xi} Application of the $\xi$-extrapolation method for the electron-electron density response $\chi_{ee}(\mathbf{q})$ of hydrogen at $\Theta=1$ and three different densities. Green crosses: direct fermionic PIMC results for $\xi=-1$; red circles: extrapolation from the sign-problem free domain of $\xi\in[0,1]$ (shaded grey area) via Eq.~(\ref{eq:fitapp}). 
}
\end{figure} 

\begin{figure}\centering
\includegraphics[width=0.48\textwidth]{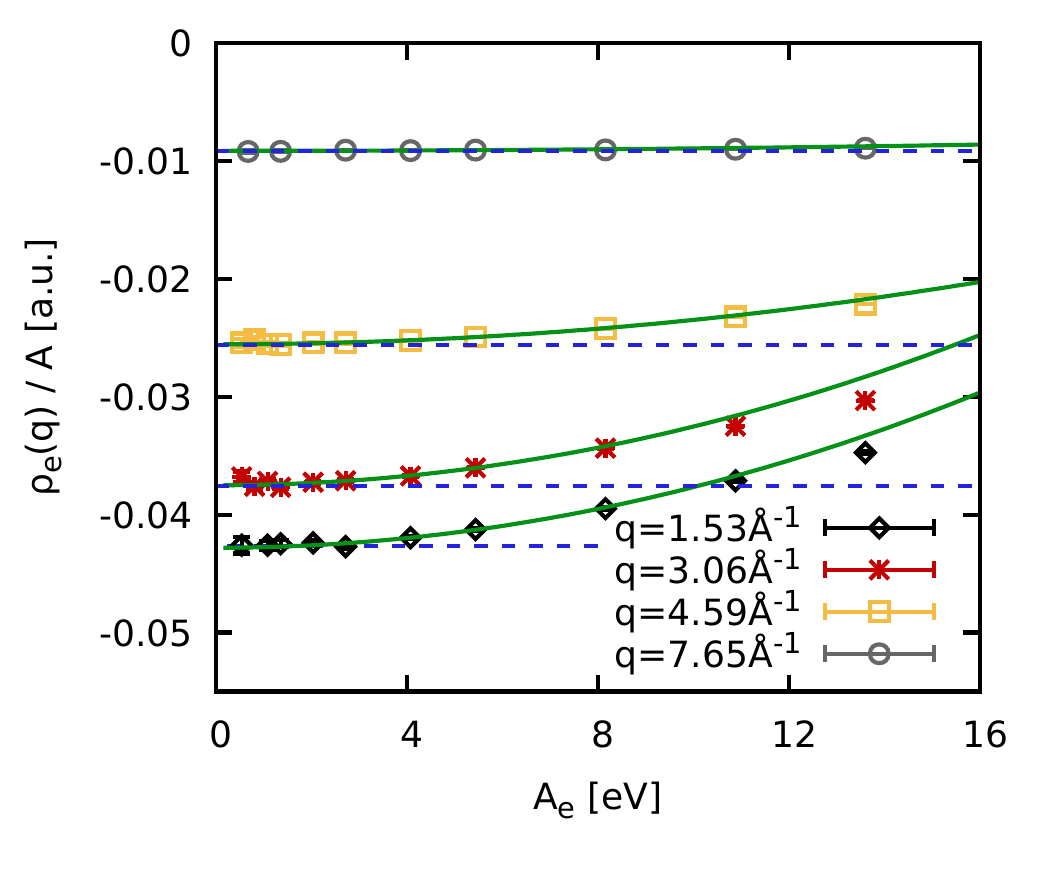}\\\vspace*{-1.3cm}\includegraphics[width=0.48\textwidth]{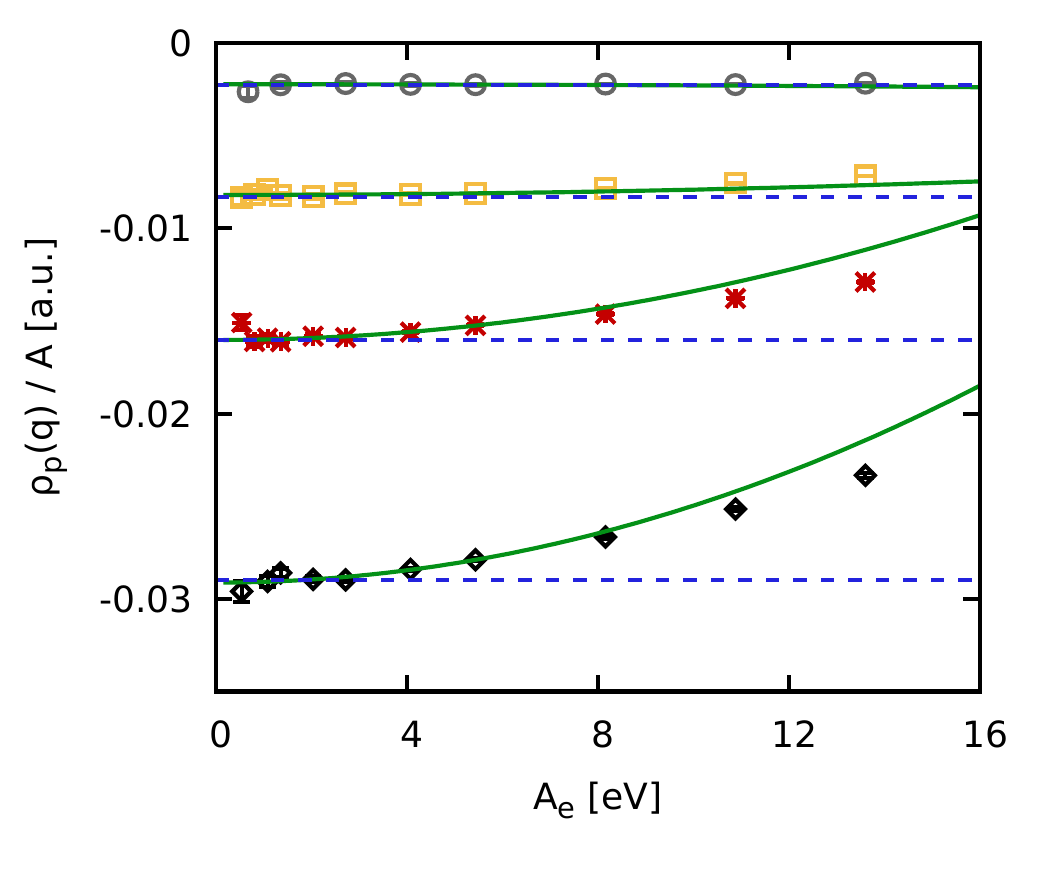}
\caption{\label{fig:appendix_direct_perturbation_rs2_theta1} Partial induced electronic $\rho_e(q)$ [top] and proton densities $\rho_p(q)$ [bottom] as a function of the electronic perturbation amplitude $A_e$ [cf.~Eq.~(\ref{eq:Hamiltonian_perturbed})] at $r_s=2$ and $\Theta=1$ for $N=14$ hydrogen atoms. Dashed blue: linear response limit evaluated from the ITCF [Eq.~(\ref{eq:static_chi})]; solid green: cubic fits via Eq.~(\ref{eq:fit}); the different symbols distinguish different perturbation wave numbers $q$.
}
\end{figure} 

With the objective to verify the absence of finite-size effects in the partial density response functions $\chi_{ab}(\mathbf{q})$ reported in the main text, we have simulated $N=32$ hydrogen atoms using the $\xi$-extrapolation method that has been originally proposed by Xiong and Xiong~\cite{Xiong_JCP_2022}, and further explored in Refs.~\cite{Dornheim_JCP_2023,Dornheim_JPCL_2024,Dornheim_Science_2024,Dornheim_HBe_2024,Xiong_PRE_2023}. Here the basic idea is to carry out a set of PIMC simulations with different values for the fictitious spin variable $\xi\in[-1,1]$, and to extrapolate from the sign-problem free domain of $\xi\geq0$ to the correct fermionic limit of $\xi=-1$ via the empirical quadratic relation
\begin{eqnarray}\label{eq:fitapp}
        A(\xi) = a_0 + a_1\xi + a_2\xi^2\,.
\end{eqnarray}
In practice, the reliability of the $\xi-$ extrapolation method can be ensured by checking its validity for a rather moderate system size, where direct fermionic PIMC simulations are still feasible.

In Fig.~\ref{fig:H_N14_rs2_theta1_xi}, we show a corresponding analysis for the electronic density response function $\chi_{ee}(\mathbf{q})$ at $\Theta=1$ for all three values of the density considered in this work. Specifically, the green crosses show the exact fermionic PIMC results for $\xi=-1$, and the red circles have been extrapolated from the sign-problem free domain (shaded grey area) via Eq.~(\ref{eq:fitapp}). We find perfect agreement between the direct PIMC result and $\xi$-extrapolated PIMC result everywhere.

\section{Direct perturbation results}\label{sec:appendix_perturbation}

In Fig.~\ref{fig:appendix_direct_perturbation_rs2_theta1}, we show additional results for the induced electron density [top] and proton density [bottom] as a function of the electronic perturbation amplitude $A_e$ (with $A_p=0$) for a variety of wavenumbers $q$. As it is expected, the induced density in the limit of $A_e\to0$ always converges towards the LRT limit (dashed blue lines) that we compute from the ITCF via Eq.~(\ref{eq:static_chi}).

\section*{Acknowledgments}
This work was partially supported by the Center for Advanced Systems Understanding (CASUS), financed by Germany’s Federal Ministry of Education and Research (BMBF) and the Saxon state government out of the State budget approved by the Saxon State Parliament. This work has received funding from the European Research Council (ERC) under the European Union’s Horizon 2022 research and innovation programme
(Grant agreement No. 101076233, "PREXTREME"). 
Views and opinions expressed are however those of the authors only and do not necessarily reflect those of the European Union or the European Research Council Executive Agency. Neither the European Union nor the granting authority can be held responsible for them. Computations were performed on a Bull Cluster at the Center for Information Services and High-Performance Computing (ZIH) at Technische Universit\"at Dresden, at the Norddeutscher Verbund f\"ur Hoch- und H\"ochstleistungsrechnen (HLRN) under grant mvp00024, and on the HoreKa supercomputer funded by the Ministry of Science, Research and the Arts Baden-W\"urttemberg and
by the Federal Ministry of Education and Research.

\bibliography{sn-bibliography}

\begin{thebibliography}{155}
\bibitem [{rep()}]{repo}%
  \BibitemOpen
  \href@noop {} {}\bibinfo {note} {A link to a repository containing all PIMC raw data will be made available upon publication.}\BibitemShut {Stop}%
\end{thebibliography}
\end{document}